\documentclass[a4paper,12pt]{article}
\usepackage[a4paper,width=160mm,top=35 mm,bottom=25mm]{geometry}
\usepackage[dvips]{graphicx}
\usepackage{color}
\usepackage{cite}
\usepackage{hyperref}
\usepackage[cmex10]{amsmath}
\usepackage{amsthm}
\usepackage{amssymb}
\usepackage{algorithmic}
\usepackage{url}
\usepackage{algorithm}
\usepackage{subfigure}
{}
{}
{}
\usepackage{subfigure}
\usepackage{float}
\usepackage{mwe}
\usepackage[detect-all]{siunitx}
\usepackage{graphicx}
\usepackage{capt-of}% or \usepackage{caption}
\usepackage{booktabs}
\usepackage{varwidth}
\usepackage{amsmath}
\usepackage{lineno,hyperref}
\usepackage{graphicx}% Include figure files
\usepackage{hyperref}
\usepackage{xcolor}
\usepackage{subfigure}
\usepackage{amssymb}
\usepackage{amsmath}
\usepackage{breqn}
\usepackage[detect-all]{siunitx}
\usepackage{dcolumn}
\usepackage{inputenc}
\usepackage{gensymb}
\usepackage{anyfontsize}
\date{}
\begin{document}
\title{On the length scale dependence of DNA conformational change under local perturbation}
\author{\normalsize Soumyadip Banerjee$^{a}$, Kushal Shah$^{b}$, Shaunak Sen$^{a}$\\ [-0.05in]
  \normalsize $^{a}$Department of Electrical Engineering\\[-0.05in]
  \normalsize Indian Institute of Technology Delhi\\[-0.05in]
  \normalsize Hauz Khas, New Delhi 110016, India\\
  \normalsize $^{b}$Department of Electrical Engineering and Computer Science\\[-0.05in] \normalsize Indian Institute of Science Education and Research (IISER)\\[-0.05in] \normalsize Bhopal 462066, Madhya Pradesh, India \\
   \normalsize Email: soumyadip@ee.iitd.ac.in$^{a}$, kushals@iiserb.ac.in$^{b}$, shaunak.sen@ee.iitd.ac.in$^{a}$         
}

\maketitle
\vspace{-3mm}
\large{\textcolor{blue}{This paper is a preprint of a paper submitted to BioSystems. If accepted, the copy of record will be available at the Biosystems online archive}}
\abstract{\noindent Conformational change of a DNA molecule is frequently observed in multiple biological processes and has been modelled using a chain of strongly coupled oscillators with a nonlinear bistable potential. While the mechanism and properties of conformational change in the model have been investigated and several reduced order models developed, the conformational dynamics as a function of the length of the oscillator chain is relatively less clear. To address this, we use a modified Lindstedt-Poincare method and numerical computations. We calculate a perturbation expansion of the frequency of the model’s nonzero modes, finding that approximating these modes with their unperturbed dynamics, as in a previous reduced order model, may not hold when the length of the DNA model increases. We investigate the conformational change to the local perturbation in models of varying lengths, finding that for the chosen input and parameters, there are two regions of DNA length in the model – first, where the minimum energy required to undergo the conformational change increases with the DNA length; and second, where it is almost independent of the length of the DNA model. We analyse the conformational change in these models by adding randomness to the local perturbation, finding that the tendency of the system to remain in a stable conformation against random perturbation decreases with increase in DNA length. These results should help to understand the role of the length of a DNA molecule in influencing its conformational dynamics.}

%\begin{keyword}
%DNA length  DNA conformational change reduced order model 
%%\MSC[2010] 00-01\sep  99-00
%\end{keyword}
%\begin{abstract}
%
%\end{abstract}
%\begin{keywords}
%DNA length\sep DNA conformational change\sep reduced order model 
%\end{keywords}
     
\section{Introduction}

A DNA sequence can manifest itself in various conformations and is observed to play an important role not only in various biological processes, such as transcription, replication, DNA repairing~\cite{Bochman,BARAT,Nouspikel}, but also in building nanostructures and nanodevices~\cite{Wilner,Simmel,Niemeyer}. The conformational state that a DNA possesses can undergo changes and is usually induced by the direct interaction of an enzyme with the DNA molecule. In fact, in an experimental study by Harada \textit{et al.}~\cite{Harada}, the enzyme RNA polymerase is observed to rotate a DNA molecule at its active site. Also, a recent review~\cite{Patel} discussed the various possible mechanisms which can lead to DNA unwinding, such as translocation and base-pair separation due to an enzyme called helicase. These studies show the importance of local interaction of enzymes in triggering such conformational changes.

A possible approach to study these conformational changes is to use a simple, coarse-grained model of a DNA. In a recent work~\cite{Peyrard,Peyrard2}, a coarse-grained model was used to quantitatively explain large localized amplitude fluctuations, called `breathing' in a DNA molecule. Also, in the work by Mezic~\cite{Mezic}, a similar coarse-grained model of a bio-molecule was considered, where the individual bases were modelled as pendula coupled to adjacent bases through torsional coupling and within a base pair through a Morse potential. A key part of his work showed local structured perturbation to trigger conformational change or flipping in the model efficiently in terms of time and energy. In addition to this, the work reported the flipping dynamics to be robust to changes in the bio-molecular size.

To explain the mechanism leading to such properties, standard tool of averaging theory~\cite{Sanders,Lichtenberg} is normally used. Recently in~\cite{Eisenhower,Eisenhower_CDC,Eisenhower_thesis}, the tool has been used to obtain a reduced order model with only a single degree of freedom . Although the reduced order model predicts an activation condition for conformational change to happen, it fails to capture the transition from one conformational state to another. A more accurate reduced order model is proposed by Du Toit \textit{et al.}~\cite{Toit}, which, unlike the one-degree model, takes into account the influence of nonzero modes in inducing transition in the flipping process. The Du Toit model approximates the nonzero modes of the full coupled oscillatory system with corresponding nonzero modes of the linear part of the full system. The approximation  	results in the coarse variable of the reduced order model to experience a time-dependent aperiodic driving.

The reduced order models mentioned work under a basic assumption that the nonzero modes influence only the zeroth or the reactive mode. However, a study of a similar system consisting of coupled non-linear oscillatory chains shows that strong resonant interaction can exist between specific nonzero modes, depending on the number of oscillators present in the system~\cite{Manevitch}. In such cases, this interaction may restrict the use of reduced order models in explaining the flipping mechanism of a coarse model of DNA molecule, especially when it has a large number of base pairs. Also, as per the recent works~\cite{Eisenhower,Toit,Koon}, the mechanism predicted by the reduced order model is seen to have a direct influence on the flipping behavior of the model subjected to local perturbation. Although the mechanism and properties of flipping are investigated in the works mentioned above, the dynamics is relatively unclear when the length of the model changes.

  In this paper, we ask about the impact of an increase in the length of the DNA strand on the applicability of a reduced order model and the flipping dynamics. We report that the assumptions required for the $1\frac{1}{2}$ degree model do not hold as the DNA length increases. We further perform a quantitative analysis of the flipping properties of the full model subjected to local perturbation. For a given parameter and input conditions, we report two different regimes of operation\textemdash first, where, for most of the range, the energy required to flip increases with the DNA length and, second, where the energy becomes almost independent to it. Finally, we add randomness to the base selection criteria and perturb the amplitude in the local perturbation process, finding that a DNA molecule is more likely to flip as the DNA length increases when perturbed randomly.

\begin{figure*}[t!]
	\centering
	\subfigure[]
	{
	\includegraphics[width=2.5in] {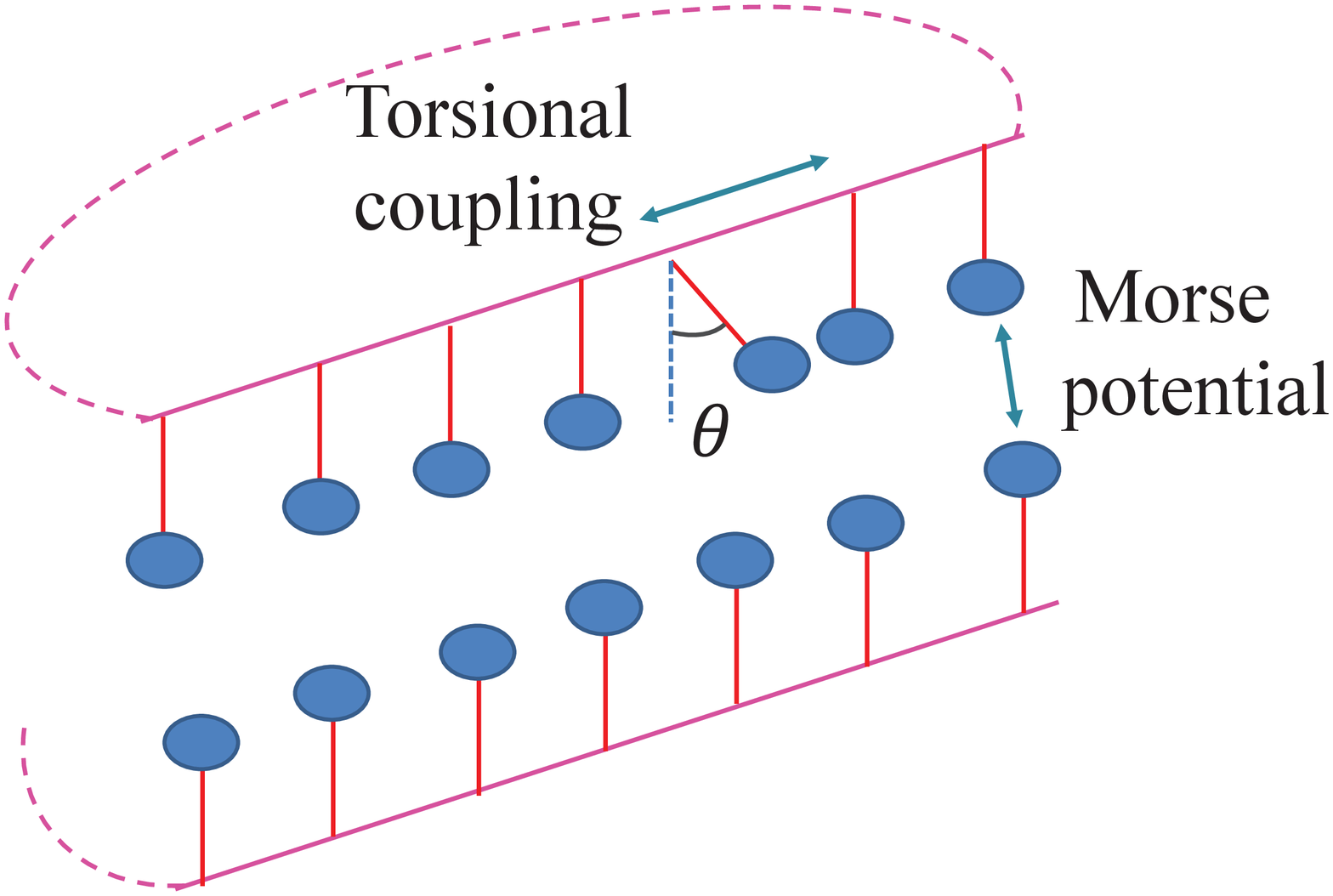}
	}
\subfigure[]
		{
\includegraphics[scale=.4]{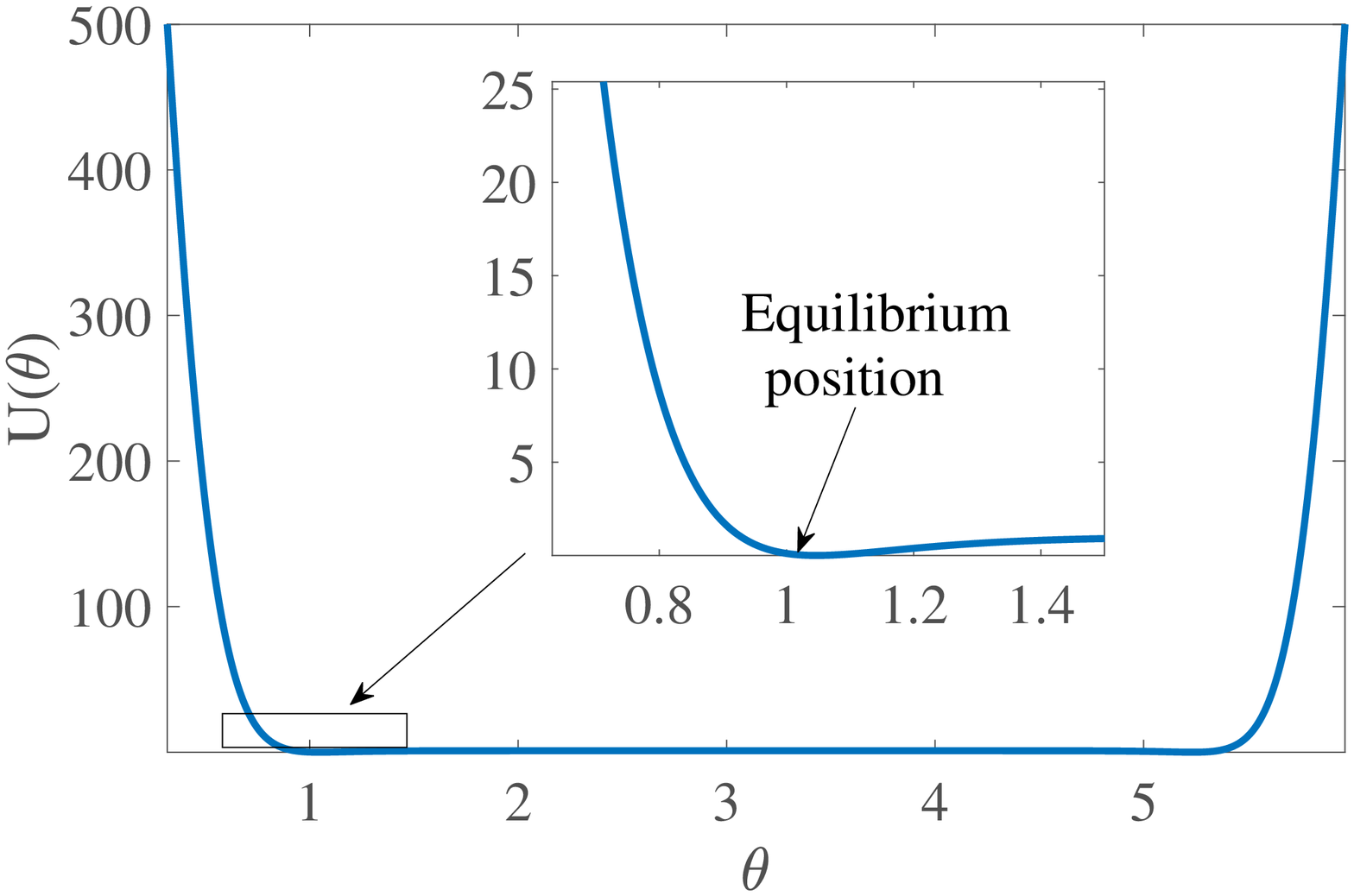}
		}
\subfigure[]
		{
\includegraphics[scale=.4]{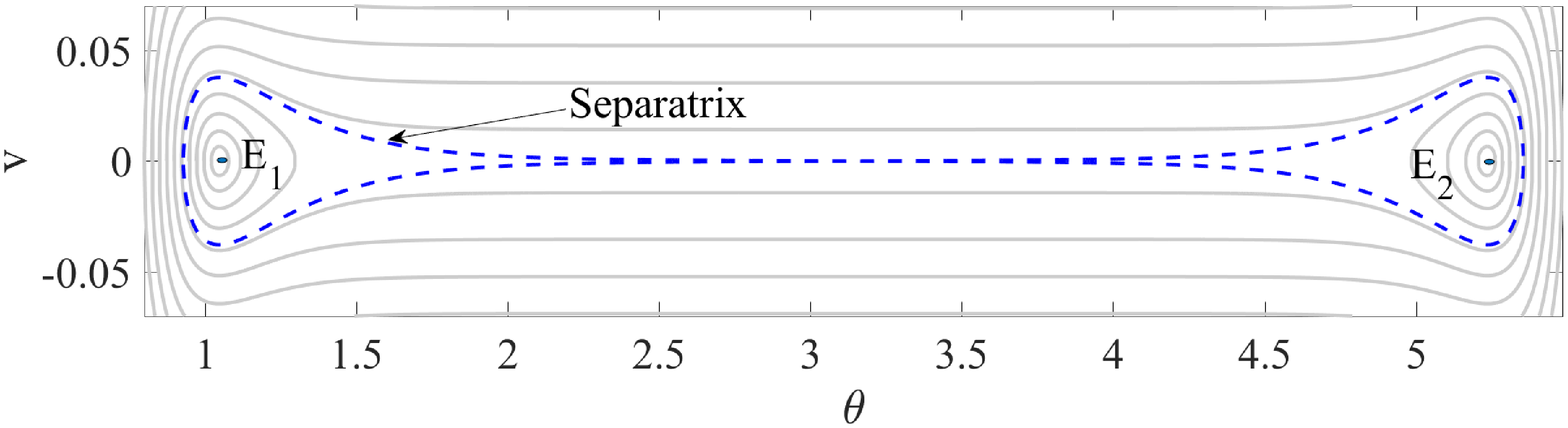}	
		}
\caption{DNA molecule modeled as a chain of coupled pendula under the influence of a bi-stable potential.\newline
\fontsize{9}{9}\selectfont{(a) Schematic showing a coarse-grained model of a DNA, where its bases are modeled as a pendula pivoted to a strand that acts as a backbone. On the same strand, the adjacent pendula are coupled through linear torsional force while the intersrand interaction between pendula on opposite strands occurs through a non-linear Morse potential. The interactions are represented with double arrows in the figure. (b) The spatial profile of the Morse potential having two stable equilibrium positions is located symmetrically about the $\theta=\pi$ axis. The typical value of the parameters used to obtain the profile are \textemdash\, $a=.7/$\AA,$h=10$\AA  and $x_{0}=5$\AA and the two stable equilibrium angular positions are given as $\theta_{0}=cos^{-1}\left(1-\frac{x_{0}}{h}\right)$=1.047 rad and $2\pi-\theta_{0}=$5.236 rad. The inset in the figure shows a magnified view of the profile near one of the equilibrium angular positions. (c) The phase portrait shows the contours of the projected Hamiltonian of the system where the angular positions are equal. The separatrix in the projected phase space, shown as a blue dashed line, separates two regions \textemdash$\,\,$one, inside the separatrix, where all the pendula oscillates around a stable equilibrium point (indicated on the figure as dots labeled as $\mathrm{E_{1}}$ and $\mathrm{E_{2}}$), exhibiting a `breathing motion' and two, outside the separatrix, where the pendula periodically crosses the $\theta=\pi$ position, exhibiting a `flipping motion'~\cite{Koon}.}
 }
\end{figure*}

\section{DNA Model}\label{DNA_model_sec}

 We first present a previously developed model of a DNA~\cite{Mezic}, where the biomolecule is approximated as a chain of coupled pendula attached to a circular strand that acts as a backbone (Figure 1(a)). The pendula represent the bases of a DNA, where the mass of each base is assumed to be concentrated at the bob of the pendulum. There are two strands in the model\,\textemdash\, first, in which the attached pendula are free to rotate on a plane orthogonal to the length of the strand and second, in which the pendula are fixed. The motion of each pendulum is governed by its interaction with two kinds of potential\,\textemdash\,first, a harmonic potential which takes into account the torsional coupling between the adjacent bases of the same strand and second, a non-linear Morse potential interaction that models bonds between opposite inter-strand bases.

This model is mathematically represented through the following equations:

\begin{equation}
\label{master_eq}
mh^{2}\ddot{\theta_{k}}-S\,\left(\theta_{k+1}-2\theta_{k}+\theta_{k-1}\right)+D\,\partial_{\theta}U\left(\theta_{k}\right)=0\,,\,\,\,\, k=1,2,....N
\end{equation}

\noindent where, $\theta_{k}$ represents the angular position of the $k^{th}$ pendulum, $N$ is the total number of pendula, $m$ is the mass of each pendulum, and $h$ is the length of the pendulum. In this model, we adopt a periodic boundary condition and, therefore, $\theta_{1}=\theta_{N+1}$ and $\theta_{0}=\theta_{N}$. The second term in (\ref{master_eq}) corresponds to the torque due to the torsional coupling, where $S$ is the torsional constant. And the third term corresponds to the torque due to Morse potential interaction, where $D$ is the Morse potential amplitude. The Morse potential with the unit amplitude is given as  $U\left(\theta\right)=\left(e^{-a\,\left(h(1-cos\theta)-x_{0}\right)}-1\right)^{2}$,\,where $a$ is the decay coefficient controlling the range over which the molecular forces between the bases act~\cite{Fidani} and $x_{0}$ is the equilibrium distance between the two nearest inter-strand bases. The non-dimensional form of the above equation can be obtained by introducing a new time-scale $\tau=\sqrt{mh^2/S}$. Under nominal parameter conditions, $S=\,42$eV , $m=\,300$AMU, $h=\,10\textrm{Å}$, the time-scale $\tau=\,$2.67 ps. For calculating time derivatives with respect to $\tau$, we have:

\begin{equation}
\label{master_eq_normal}
\ddot{\theta_{k}}-\left(\theta_{k+1}-2\theta_{k}+\theta_{k-1}\right)+\epsilon\,\partial_{\theta}U\left(\theta_{k}\right)=0\,,
\end{equation}

\noindent where, $\epsilon=\frac{D}{S}$ denotes the relative amplitude of the Morse potential with respect to the harmonic torsional potential. Since the hydrogen bonds between the base pairs of the complementary strands are weak as compared to the covalent bonds that make up the strands~\cite{Sinden}, the dimensionless parameter $\epsilon\ll1$. For the given parametric regime, the Morse potential has two global minima located symmetrically about $\theta=\pi$ rad, as shown in Figure 1(b). As the model represent a conservative system, the total energy of the system is a constant of motion, given by the Hamiltonian, $E$, 

\begin{equation}
\label{Hamiltonian}
E=\sum_{k=1}^{N}\frac{\dot{\theta}_{k}^{2}}{2}+\frac{1}{2}\left(\theta_{k}-\theta_{k-1}\right)^{2}+\epsilon\,U\left(\theta_{k}\right),
\end{equation}
			
A structural property of the above described model is that it has two conformational states, wherein each, all pendula are equally displaced from their complementary pendula by the equilibrium distance, $x_{0}$. The states correspond to two stable equilibrium points located symmetrically in the phase space of the system. Figure 1(c) shows the location of the points in a projected phase portrait of the system where the angular positions of the pendula are equal. A system in one of these conformed states can be subjected to a local perturbation~\cite{Mezic}, where a single or a group of spatially close pendula are disturbed from their resting positions. Under such perturbation, the trajectory of the system in phase space can either remain close to the stable equilibrium point or transit to a region near the other equilibrium point.

\section{Flipping dynamics}\label{Flipping_dynamics_sec}
 We associate two characteristic features to the local perturbation process\,\textemdash\, the number of targeted pendula to be perturbed and the initial energy imparted to the system, termed as `perturbation energy'. We consider an example of such a perturbation through the following set of initial conditions:

\begin{equation}
\label{IC}
\begin{split}
 \left(\theta_{k}(0),\dot{\theta}_{k}(0)\right)&=\left(\theta_{i},0\right)\quad k=[N/2, N/2+1],\\&=(\theta_{0},0)\quad k\neq [N/2, N/2+1],
 \end{split}
\end{equation}

\noindent where, $\theta_{0}$ is the equilibrium angular position.
\noindent Physically, the initial conditions denote that the two pendula at the centre are pushed into the repulsive region where the new angular position is $\theta_{i}$, while the other pendula are located at $\theta_{0}$. Note that, as the above model has a translational invariance, we can choose the target pendula from any arbitrary location. 

Using the above initial condition and the parametric condition, $a=0.7$, $h=10$, $x_{0}=5$ and $\epsilon=1/1400$, we simulate$\,\left(\ref{master_eq_normal}\right)$ for 30 pendula~$\left(N=30\right)$, for a duration of 2000 units with a time step of 0.01 units. Here, as well as in the rest of the paper, we use the fourth-order Runge-Kutta method~\cite{Chapra} to perform the simulation. We find that the method conserves energy relatively well during the simulation (Figure 2(c)).
\begin{figure*}[t!]
	\centering
	\subfigure[]
	{
	\includegraphics[width=5.5in] {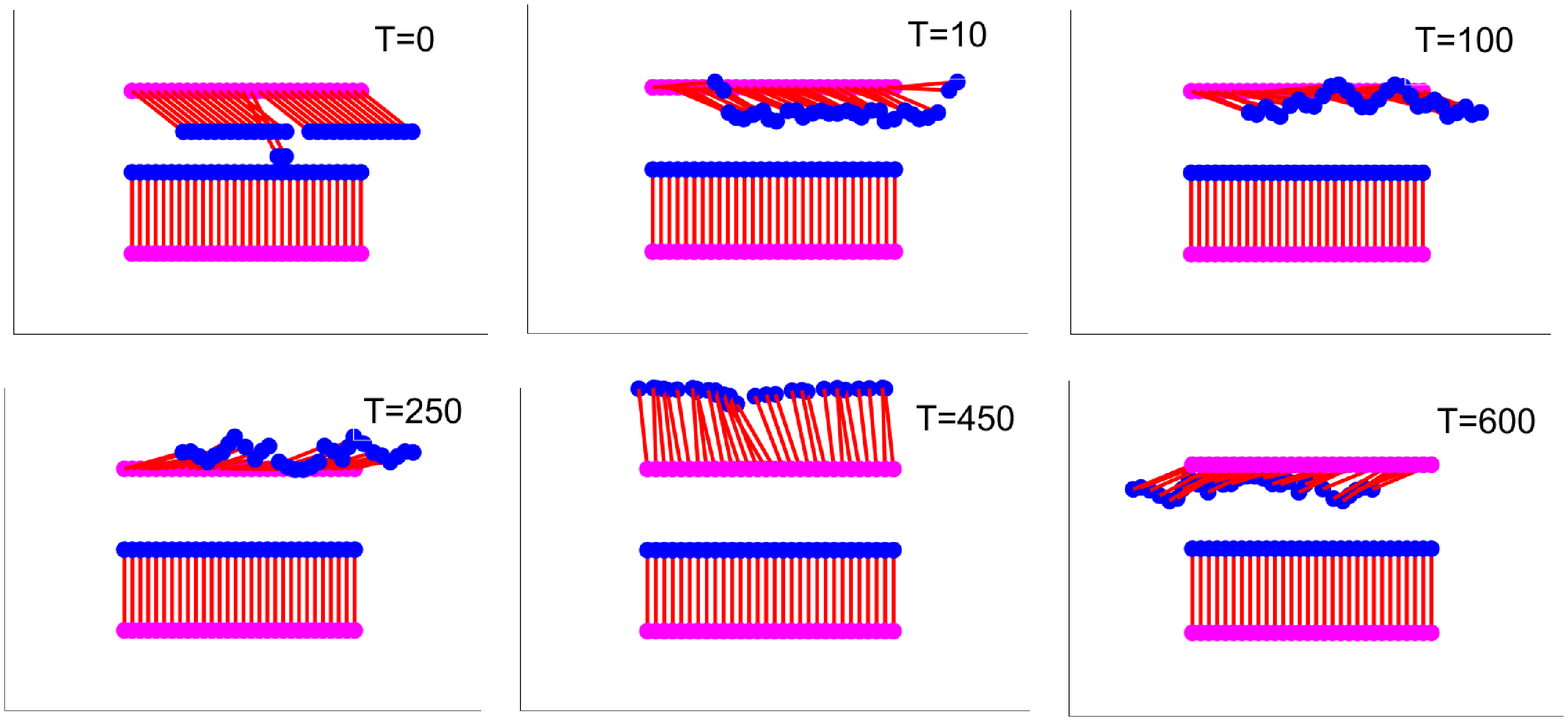}
	
	}
	
\subfigure[]
		{
\includegraphics[scale=.24]{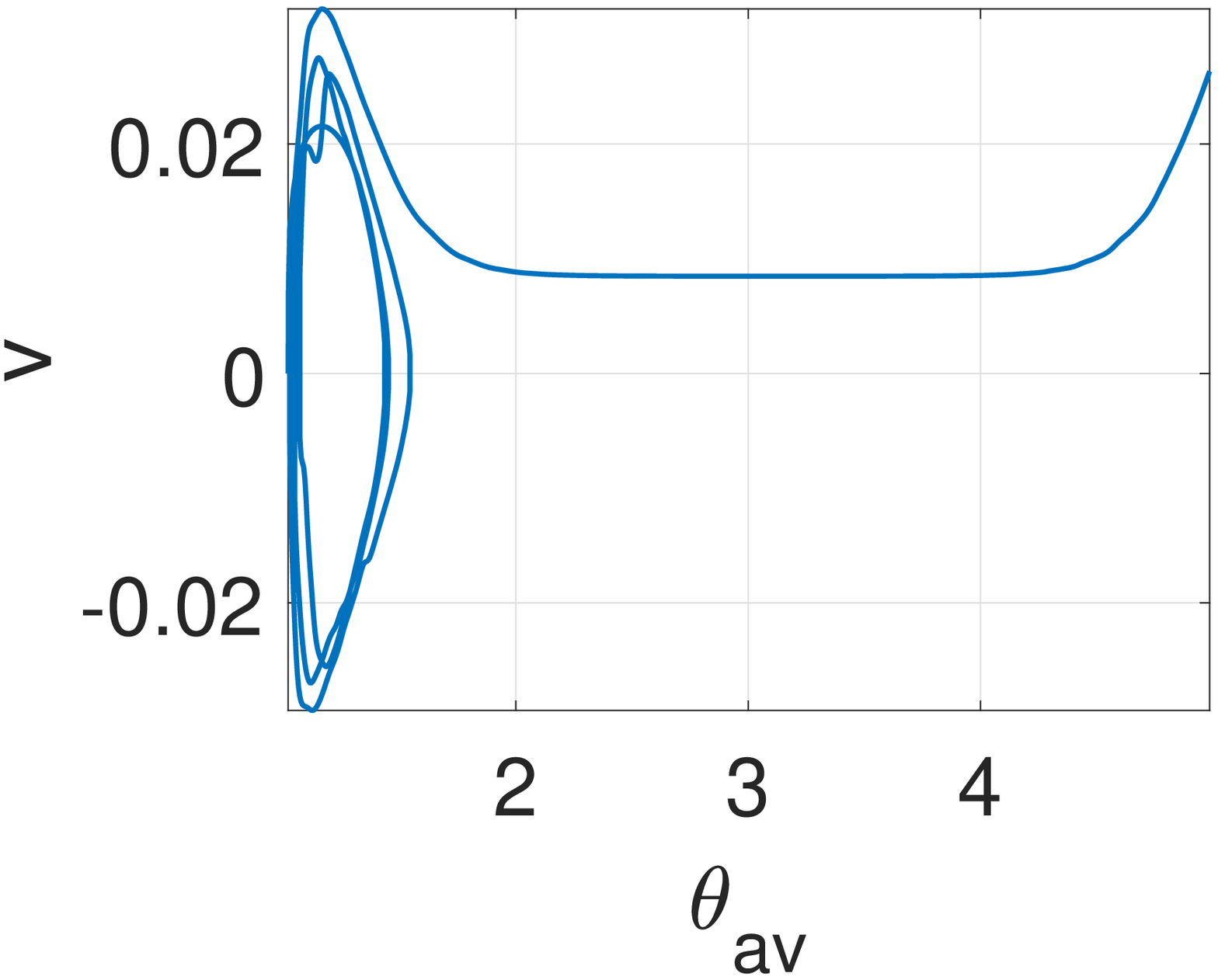}
		
		}
		\label{model_sim2}
\subfigure[]
		{
\includegraphics[scale=.24]{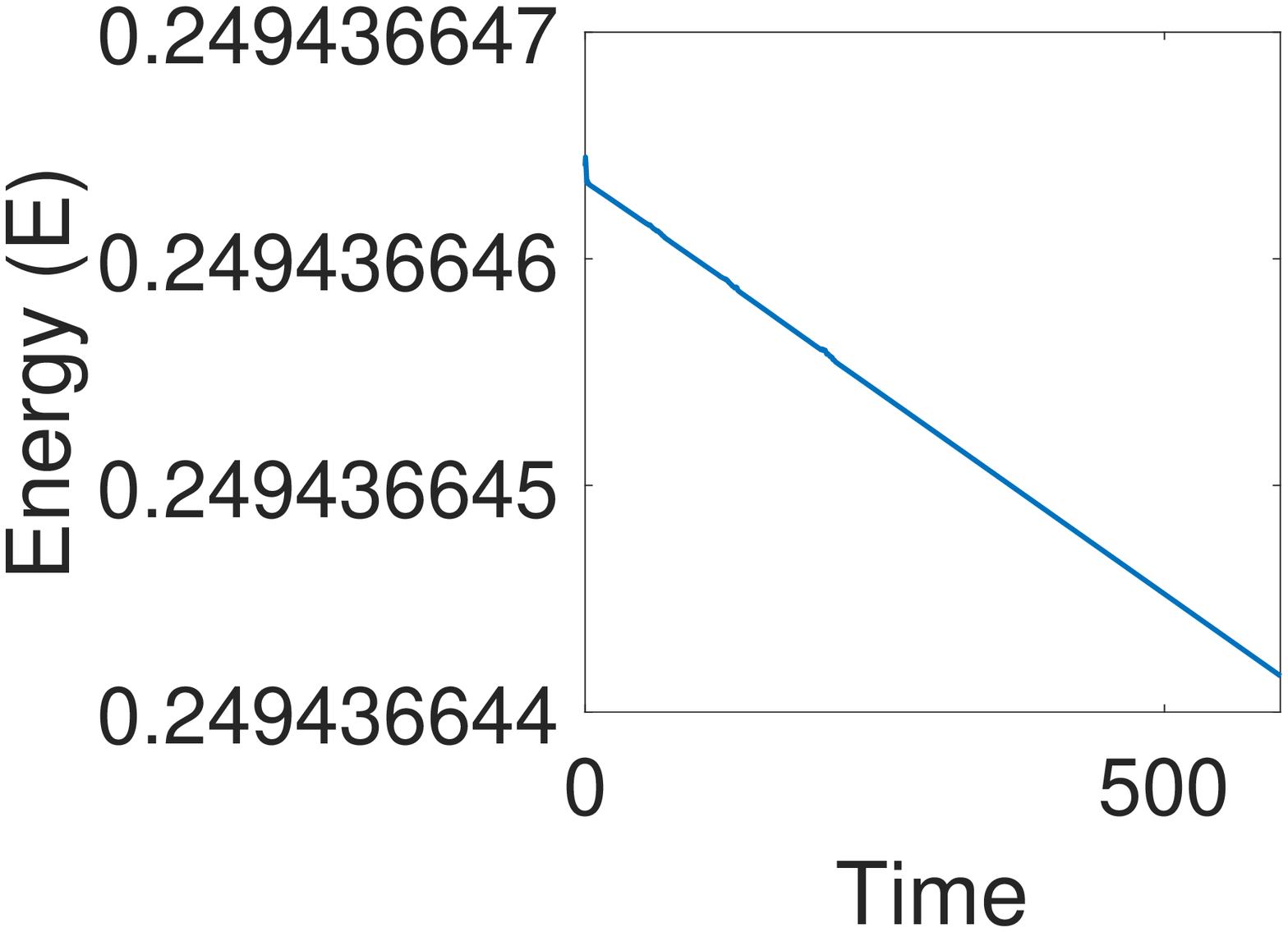}	
		}
		\label{model_sim2}		
		\subfigure[]
		{
\includegraphics[scale=.235]{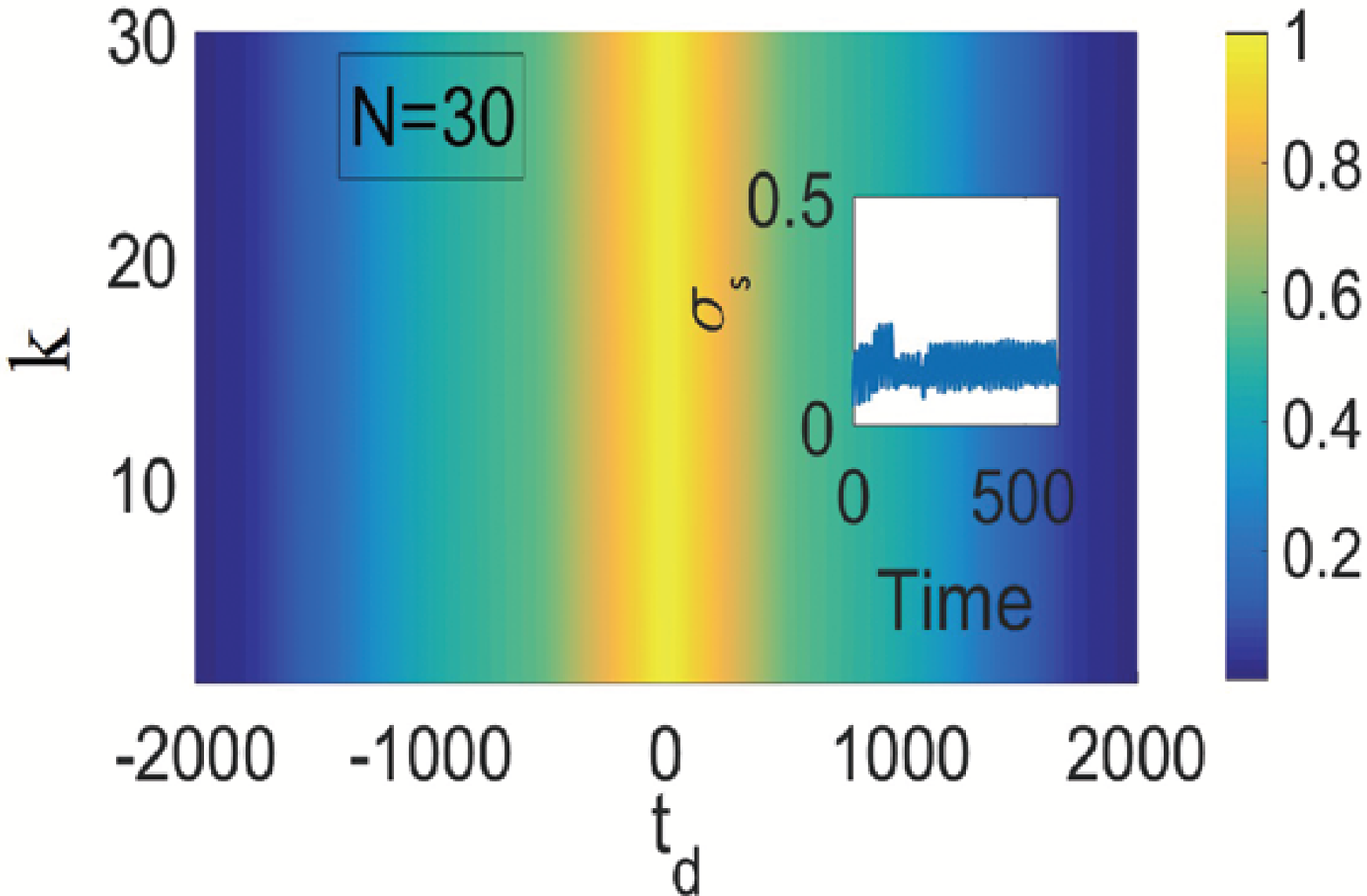}
		}
	\label{model_sim3}
	
	\caption{Flip dynamics of the DNA model under the effect of a local perturbation. \newline\fontsize{9}{9}\selectfont{In the local perturbation process, the two pendula in the middle (at $\frac{N}{2}$ and $\frac{N}{2}+1$) are shifted to a new angular position $\theta_{i}=0.635$ rad  while the rest of the pendula are kept at their equilibrium positions $\theta_{c}=1.047$ rad. (a) Snapshots showing the time evolution of the DNA configuration from one stable equilibrium zone to the other, obtained by simulating $\left(\ref{master_eq_normal}\right)$. (b) Phase portrait showing the trajectory of the DNA system projected onto the average co-ordinates ($\theta_{av}$,$v$). (c) Figure showing the total energy of the system as a function of time. (d) Plot of the cross-correlation coefficient $\rho$ as a function of  $t_{d}$ and the pendulum number $k$. The color bar indicates the value of $\rho$. The inset shows the standard deviation $\sigma_{s}$ of the angular position of a pendulum from the average position of all the pendula during the simulation.}
	}
\end{figure*}

The DNA dynamics under the aforementioned local disturbance is as follows: Initially, the position of the two pendula at the centre is disturbed, through which energy is imparted into the system in the form of potential energy at the time, $T=0$. The perturbation energy provided to the targeted pendulums eventually spreads to all the other pendula on both sides, thereby perturbing them from their resting positions. In the initial phase, the pendula move in a correlated manner near the stable equilibrium position. In this phase, the angular positions of the pendula remain bounded and the system is said to be in a `breathing state'~\cite{Chou}, where the base pairs of a DNA molecule undergo a closed-open motion. Figure 2(a) shows snapshots of the system's configuration at two such time instants ( $T=\,$10 units and $T=\,$100 units\,), where the position of the pendula are visibly bounded near the stable equilibrium point. After a certain period, a second phase is observed where the angular positions of the pendula start to increase. The fourth and fifth snapshots( $T=\,$250 and $T=\,$450) capture moments of this coherent rise in the angular positions of the pendula. It is in this phase where all the pendula escape from near one stable equilibrium point and collectively move towards the another stable equilibrium point (its snapshot is shown at time $T=\,$600).

The transition from a bounded to an unbounded motion can be tracked using the average angular position of the pendula, which is calculated as:

\begin{equation}\label{average_angle}
\theta_{av}=\frac{1}{N}\sum_{k=1}^{N}\theta_{k}\,,
\end{equation}

\noindent and the corresponding average angular velocity is denoted by $v$. Figure 2(b) shows the average angular trajectory of the system following the contour of its projected Hamiltonian. The contours pass through a region near the stable equilibrium point ($\theta_{av}=\theta_{0}$ and $v=0$) in the phase space of the average co-ordinates called the resonance zone~\cite{Mezic}, where the system has relatively strong interaction with the Morse potential. Note that the collective transition of the pendula is detected when the average angular position crosses the $\pi$ mark. In the rest of the paper, this event and the time it takes for it to happen are referred to as a `flip' and `flip time', respectively.

To understand the collective behavior of the pendula, we calculate the cross-correlation coefficient, $\rho$ between the time-series data of the angular position of the $\frac{N}{2}^{th}$ pendulum and that of the $k^{th}$ pendulum, given as,
\begin{equation}\label{correlation}
\rho(k,t_{d})=\frac{\int_{0}^{T_{0}}\theta_{\frac{N}{2}}(t)\,\theta_{k}(t-t_{d})\,dt}{\sqrt{\int_{0}^{T_{0}} \theta_{\frac{N}{2}}^{2}(t)\,dt\;\int_{0}^{T_{0}}\theta_{k}^{2}(t)\,dt}}\,,
\end{equation}
\noindent where, $t_{d}$ is the displacement time and $T_{0}$ is the total simulation run time. Note that the restriction over the integral limits is because the time-series data of the angular position considered are time-limited between $T=0$ and $T=2000$ units.  Here, $\rho$ is calculated numerically in Matlab using a built-in function called \textit{xcorr}, with `coeff' as the normalization option. From the plot shown in Figure 2(d), we find that the motions of the pendula are highly correlated at $t_{d}=0$. Also, we find the $\rho$ value to change with displacement time, $t_{d}$. Due to edge effect at the beginning and end of the signals, the interpretation of correlation at larger displacement time may be inaccurate. Nevertheless, we restrict our observation within the displacement time range $\pm1000$ units as they ensure at least 50 $\%$ overlap of the signals during the computation of $\rho$.   

In the above example, the collective motion of the system is such that the angular position of each pendulum remains in the vicinity of the average angular position of all the pendula with a standard deviation of $\sigma_{s}\approx0.2$ (please see inset of Figure 2(d)). In fact, during the rotation or the flipping period, the motion of the system resembles a rigid body motion~\cite{Goldstein} where the angular distance between any two pendula remains almost equal. These rigid motions can be efficiently triggered by low-frequency modes of the system~\cite{Eisenhower}. In fact, in a recent work~\cite{Delarue} that uses normal mode analysis, low-frequency modes have also been found to effectively induce large amplitude collective motion of atoms in a biomolecule.

The mechanism behind the flipping behavior described above can be explained through the dynamics of the Fourier modes~\cite{Eisenhower}. These modes are obtained by projecting the spatial angular co-ordinates on to the Fourier space through the following co-ordinate transformation:

\begin{equation}\label{transform}
\Theta=T\,\bar{\Theta},
\end{equation}

\noindent where, $\Theta=[\theta_{1}\;\theta_{2}.............\theta_{N}]'$  is the set of angular variables in the real space and $\bar{\Theta}=[\bar{\theta}_{0}\;\bar{\theta}_{1}.............\bar{\theta}_{N-1}]'$ is the set of modal co-ordinates in the Fourier space. $T$ is a real symmetric orthonormal matrix which relates the two spaces~\cite{Koon} and its columns are the eigenvectors in the configuration space. Note that $\bar{\theta_{0}}$ is proportional to the average angular position of the pendula, $\bar{\theta}_{0}=\sqrt{N}\theta_{av}$. Using $\left(\ref{master_eq_normal}\right)$ and $\left(\ref{transform}\right)$, the equation for the modal co-ordinates can be obtained as
\begin{equation}\label{modal_eq}
\ddot{\bar{\theta}}_{w}=-\alpha_{w}^{2}\,\bar{\theta}_{w}-\epsilon\sum_{n=1}^{N}T_{nw}G\left(\sum_{w'=0}^{N-1}T_{nw'}\bar{\theta}_{w'}\right),
\end{equation}

\noindent where $w$ is the mode number, $T_{nw}$ represents the $n^{th}$ component of the eigenvector corresponding to mode $w$ as defined in the configuration space of the unperturbed system, $G(\theta)=\partial_{\theta}U$ and $\alpha_{w}=2sin\left(\frac{\pi w}{N}\right)$ correspond to the eigenvalue of the unperturbed system’s state matrix, which is obtained when $\epsilon$ is set to zero in  $\left(\ref{modal_eq}\right)$. In the unperturbed model, the modes are the decoupled simple harmonic oscillators which are free to oscillate at their characteristic frequencies, $\alpha_{w}$. However, in the perturbed model, the modes become coupled and are allowed to exchange energy among themselves. The total energy of the system in the coupled condition can be given in terms of the modal co-ordinates as:

\begin{equation}\label{Hamiltonian_modal_space}
H=\sum_{w=0}^{N-1}\left(\frac{\bar{p}_{w}^{2}}{2}+\frac{1}{2}\alpha_{w}^{2}\bar{\theta}_{w}^{2}\right)+\epsilon\sum_{n=1}^{N}U\left(\sum_{w=0}^{N-1}T_{nw}\bar{\theta}_{w}\right),
\end{equation}

\noindent where the first term of the Hamiltonian is the summation of the modal energies and the second term, with $\epsilon$ as a factor, represents the energy associated with the interacting modes. In a coupled-oscillatory system, such as the one studied here, the modes can interact with each other through their internal resonance~\cite{manevitch_book}. A condition for it is given as~\cite{Eisenhower}:

\begin{equation}\label{resonance_cond}
   \left(\overrightarrow{k}.\overrightarrow{\alpha}\right)<\frac{1}{c|\overrightarrow{k}|^{v}},\,\,|\overrightarrow{k}|=|k_{0}|+|k_{1}|+.....+|k_{N-1}|,
\end{equation}

\noindent where, $\overrightarrow{k}.\overrightarrow{\alpha}=k_{0}\alpha_{0}+k_{1}\alpha_{1}+...+k_{N-1}\alpha_{N-1},\,\,k\in\mathbb{Z}^{N}$\textendash\,[0] and $c$ and $v$ are constants. The inequality is satisfied if the left-hand side is equal to zero. This can happen if the modal frequencies are commensurable. However, in certain cases, internal resonance can occur among incommensurate frequencies. An example of such an occurrence can be found in~\cite{Cartwright}, where the universality of three-frequency resonance is demonstrated. The right-hand side of the$\left(\ref{resonance_cond}\right)$ takes into account the resonance zone in the phase space, where the modal frequencies are nearly commensurate~\cite{Eisenhower_thesis}. An example of such resonance is the `nearly 0:1 resonance'~\cite{Koon}, where the zeroth order mode resonates with all the nonzero order modes of the system. Nonlinear interaction within the system can also perturb the frequencies of the modes having incommensurable frequencies and make them commensurable for resonance to take place. The resonance condition in such cases may exist for a finite duration, as it will depend on how long the system stays close to the resonance zone~\cite{Eisenhower_thesis}.

\section{Validating reduced order model}\label{Validating_reduced_order_model}
 To understand the influence of resonance in such systems, we further analyze the modal dynamics of the coupled pendulum system studied in the previous section. By tracking the time evolution of the average angle and the energy of the selected modes (Figure 3(a)\,\textemdash\,(b)), we find that the change in the modal energies depends on how close the average mode is to the resonance zone. A discrete change in the energy level of some modes is observed whenever the system comes relatively closer to the stable  equilibrium position, indicating resonance between the modes. However, when the system is relatively far from the equilibrium point, the energy in the nonzero mode remains almost constant and the system approximately follows an integrable motion~\cite{wiggins} which corresponds to the nonzero modes, represented as:

\begin{equation}\label{mode_sol}
\bar{\theta}_{w}^{approx}=A_{w}\,cos\,\alpha_{w}t+\frac{B_{w}}{\alpha_{w}}\,sin\,\alpha_{w}t=\sqrt{\frac{2E_{w}}{\alpha_{w}}}cos(\phi_{w}),\quad\quad w=1,2\cdot\cdot N-1
\end{equation}  
\begin{figure*}[t!]
	\subfigure[]
	{
	\includegraphics[width=3in] {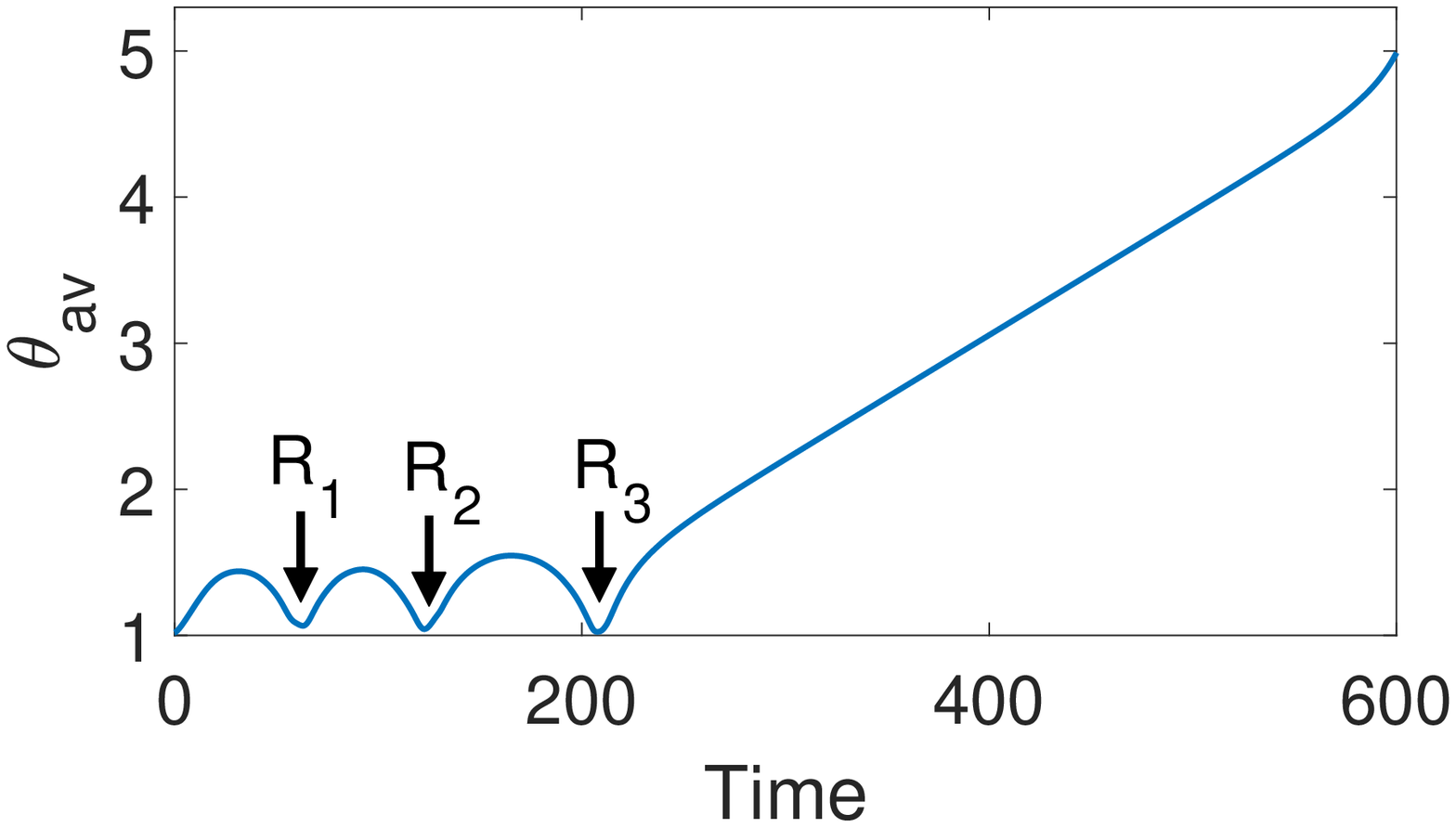}
	}~
	\label{timetrace_avg}
	\subfigure[]
	{
	\includegraphics[width=3in] {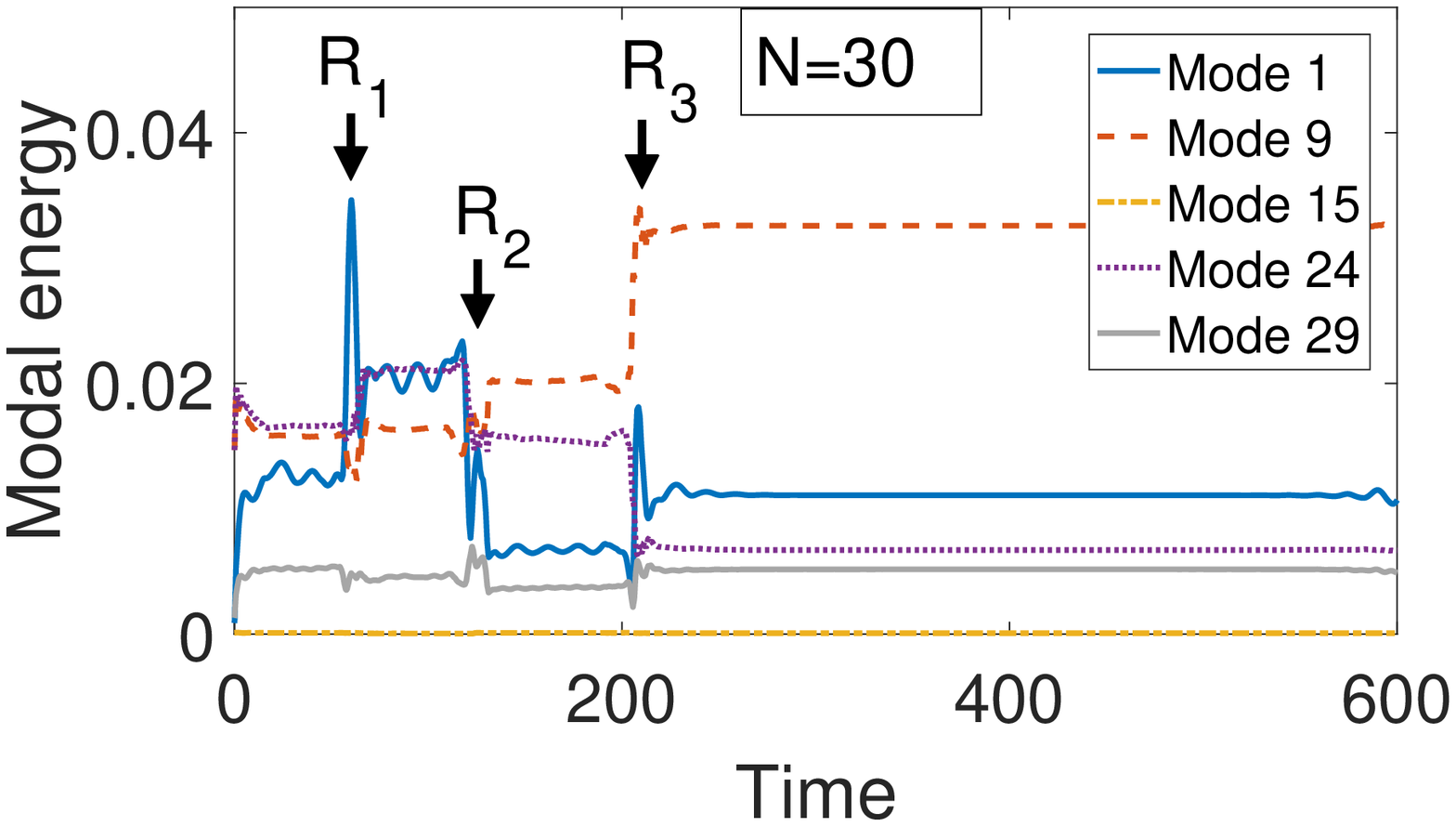}
	}
	\label{mode_energy}
	
\caption{Resonance between the modes of the model can contribute to flipping in the DNA model. \newline\fontsize{9}{9}\selectfont{(a) Figure shows the time trace of the average angular positions of the pendula. The arrows in the figure indicate the point of closest approach to the equilibrium position, labelled as $\mathrm{R_{1}}$, $\mathrm{R_{2}}$ and $\mathrm{R_{3}}$.(b) Figure shows the time evolution of the modal energies of the selected modes of the system.}}
\end{figure*}\noindent where, the modal phase, $\phi_{w}=\alpha_{w}t+\psi_{w}$ and $A_{w}$ and $B_{w}$ are constants which depend on the initial conditions, $E_{w}$ is the energy in the $w^{th}$ mode and $\psi_{w}=-tan^{-1}\left(\frac{B_{w}}{\alpha_{w}A_{w}}\right)$. The above approximation is used to obtain a reduced order model in~\cite{Toit}. In this model, it is assumed that from the onset of perturbation, the nonzero modes follow unperturbed dynamics. In such cases, the only mechanism triggering the flip event is the driving of the zeroth order mode by the unperturbed nonzero modes, as such interaction would not perturb the dynamics of the nonzero modes. However, if the trajectory of the system in phase space passes through any resonance region, the energy associated with the nonzero modes can get largely perturbed, in which case reducing the order of the full model by averaging techniques may not be possible~\cite{Neishtadt}.

A basic assumption of the reduced order model is that the frequency of the perturbed modes is equal to that of the unperturbed modes. Although the assumption holds outside the resonance zone~\cite{Eisenhower}, it may not hold inside it. A possible region to test the assumption would be a region close to a stable equilibrium point. Previously, in~\cite{Koon}, an approximate expression of the angular frequency of the perturbed mode was obtained. However, it used the partial averaging method which works outside the resonance zone~\cite{Eisenhower}. To compute the perturbed frequency close to the equilibrium point, we use the modified Lindstedt-Poincare method~\cite{He}, as it takes the internal resonance into account by inherently eliminating its secular terms.

Following procedure outlined in~\cite{He}, we expand the modal co-ordinate of the $w^{th}$ mode, $\bar{\theta}_{w}$ and the unperturbed modal frequency $\left(\alpha_{w}\right)$ in $\left(\ref{modal_eq}\right)$ in different orders of $\epsilon$,

\begin{equation}
\bar{\theta}_{w}=\bar{\theta}_{0,w}+\epsilon\,\bar{\theta}_{1,w}+\cdot\cdot\cdot,
\end{equation}
\begin{equation}\label{freq_expand}
\alpha_{w}^{2}=\Omega_{w}^{2}+\epsilon\,\Omega_{w,1}^{2}+\cdot\cdot\cdot,
\end{equation}

\noindent where $\bar{\theta}_{0,w}$, $\bar{\theta}_{1,w}\cdot\cdot\cdot$ are the correction terms, $\Omega_{w}$ is the perturbed angular frequency of the mode `$w$', and $\Omega_{w,1}$, $\Omega_{w,2}\cdot\cdot\cdot$  are terms chosen to eliminate secular terms. Collecting different orders of $\epsilon$, in $\left(\ref{modal_eq}\right)$, we get,

\begin{equation}
\label{mode_eq_order}
\begin{split}\epsilon^{0}\, & :\qquad\qquad\qquad\ddot{\bar{\theta}}_{0,w}+\Omega_{w}^{2}\,\bar{\theta}_{0,w}=0,\\
\epsilon^{1}\, & :\qquad\qquad\qquad\ddot{\bar{\theta}}_{1,w}+\Omega_{w}^{2}\,\bar{\theta}_{1,w}=-\Omega_{w,1}^{2}\,\bar{\theta}_{0,w}\\
& \qquad\qquad\qquad\qquad\qquad\qquad-\sum_{n=1}^{N}T_{nw}G\left(\frac{1}{\sqrt{N}}\bar{\theta}_{0,0}+\sum_{w'}T_{nw'}\bar{\theta}_{0,w'}\right),
\end{split}
\end{equation}

\noindent Next, we perturb the system initially placed at the equilibrium position by providing $E_{w}$ amount of energy to the $w^{th}$ nonzero mode. If the energy provided to the mode is such that $ah\sqrt{\frac{E_{w}}{N}}<1$, the expression of the perturbed modal frequency can be written as (please see appendix for details):

\begin{equation}\label{pert_freq}
\Omega_{w}^{2}=\alpha_{w}^{2}+\epsilon\,2a^{2}x_{0}\left(2h-x_{0}\right)+\epsilon\,\mathcal{O}\left(a^{2}h^{2}\frac{E_{w}}{N}\right),
\end{equation}

\noindent Under the perturbing condition considered here, the perturbed modal frequency differs from the unperturbed modal frequency by an error that is approximately represented by the second term. For instance, if the energy provided to the system is arbitrarily small, the second term dominates over the other error terms. 
Further, it can be seen that the reduced order model is valid if the first term on the right is much larger than the rest of the terms. For the given case, the reduced order model may not hold if, 
\begin{equation}
\alpha_{1}^{2}<<2\epsilon a^{2}x_{0}\left(2h-x_{0}\right)+\epsilon\,\mathcal{O}\left(a^{2}h^{2}\frac{E_{w}}{N}\right),
\end{equation}

\noindent Here, the frequency corresponding to the first mode, $\alpha_{1}$, is chosen as it is the minimum possible characteristic frequency of the unperturbed system. On rewriting the above equation and assuming $\alpha_{1}=2sin\left(\frac{\pi}{N}\right)\approx\frac{2\pi}{N}$, we have,

\begin{equation}
N>>\sqrt{\frac{4\pi^{2}}{2\epsilon a^{2}x_{0}\left(2h-x_{0}\right)+\epsilon\,\mathcal{O}\left(a^{2}h^{2}\frac{E_{w}}{N}\right)}},
\end{equation}

\begin{equation}\label{N_inequal}
\Rightarrow N>>\sqrt{\frac{4\pi^{2}}{2\epsilon a^{2}x_{0}\left(2h-x_{0}\right)}}\approx28,
\end{equation}

\noindent The above inequality $\left(\ref{N_inequal}\right)$, with the right-hand side being the limiting value, may put a restriction on the DNA length for the reduced order model to hold.
\begin{figure*}[t!]
	\centering
	\subfigure
	{
	\includegraphics[width=2.1in] {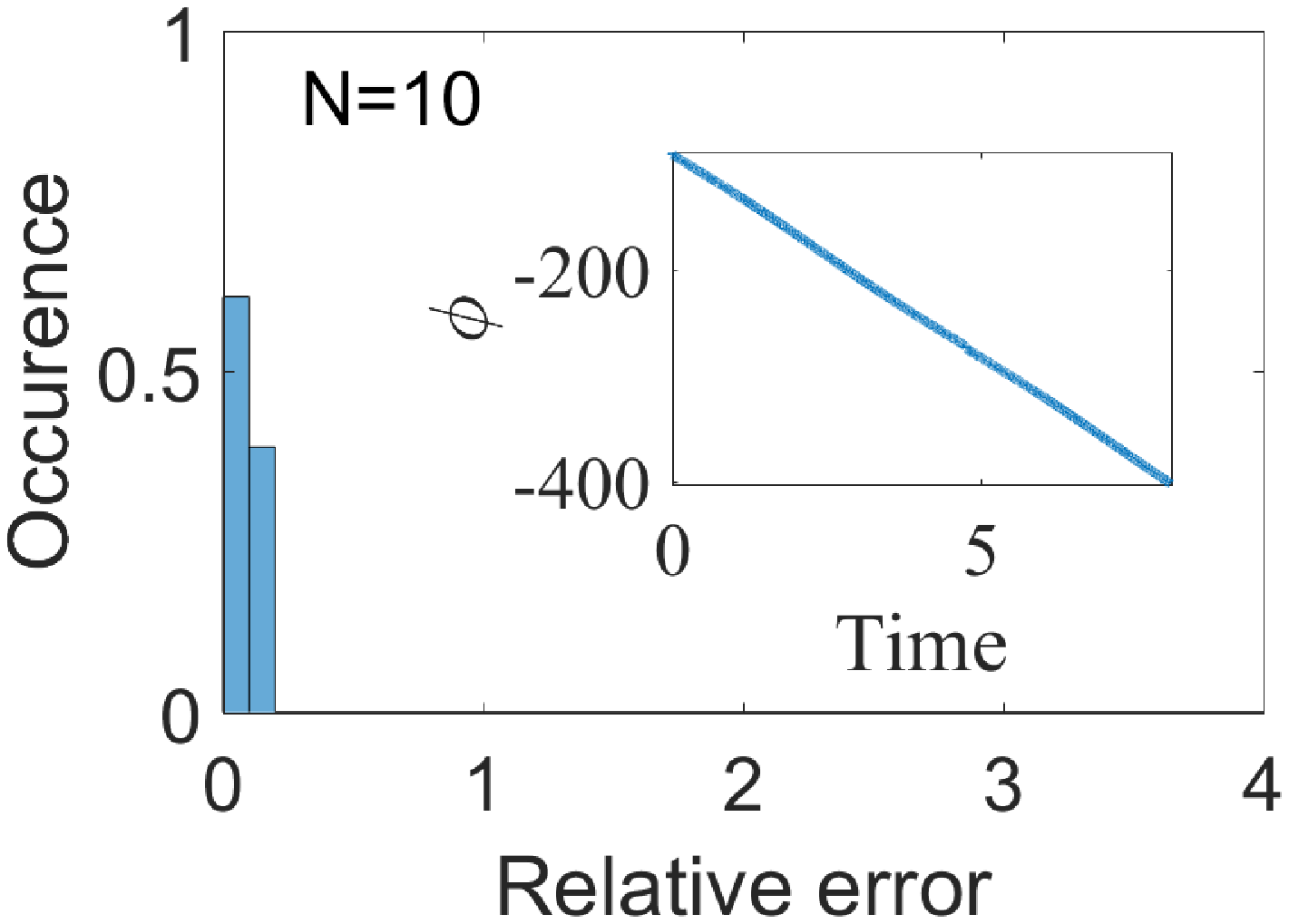}
	}~
	\subfigure
	{
	\includegraphics[width=2.1in] {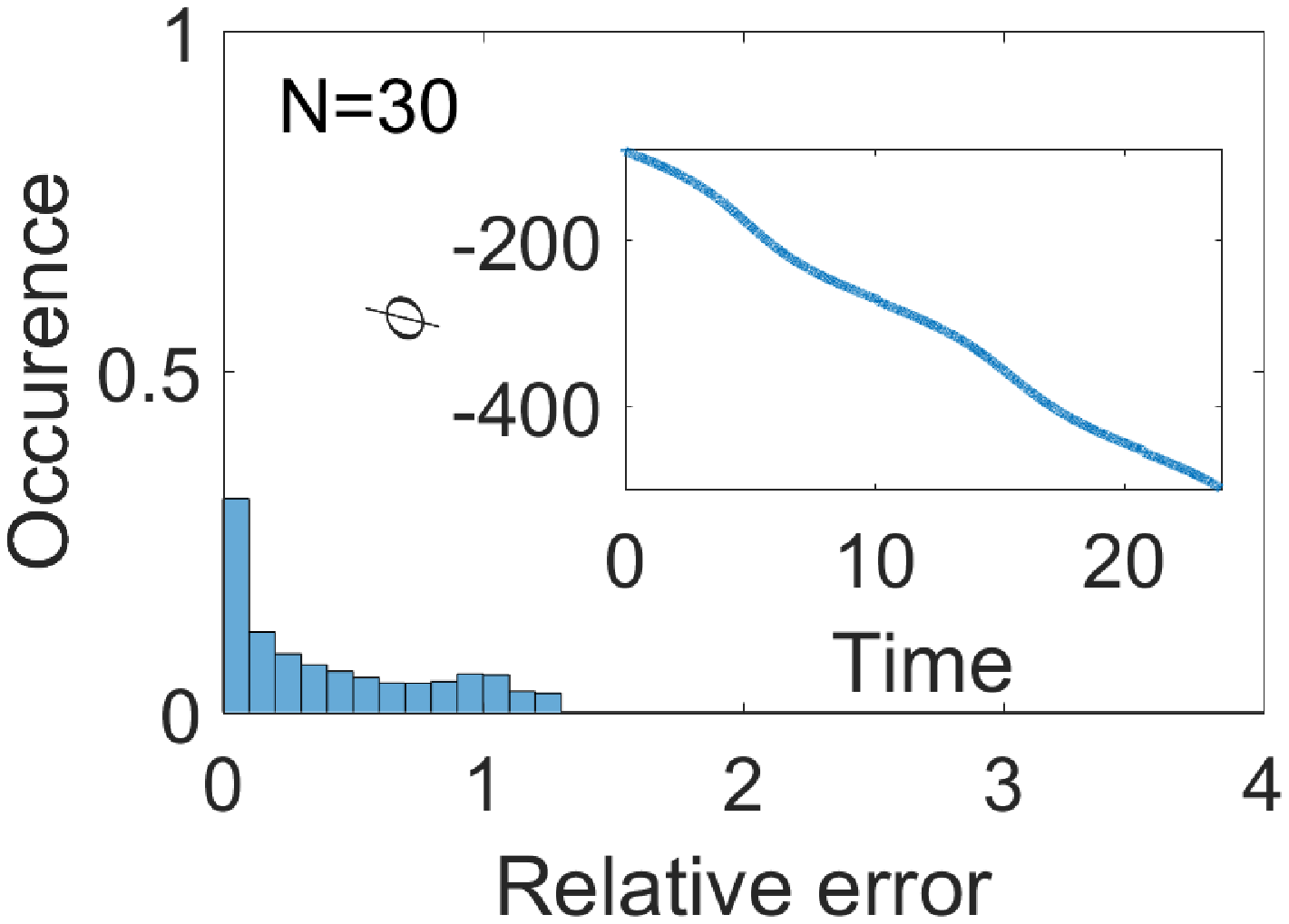}
	}~
	\subfigure
	{
	\includegraphics[width=2.1in] {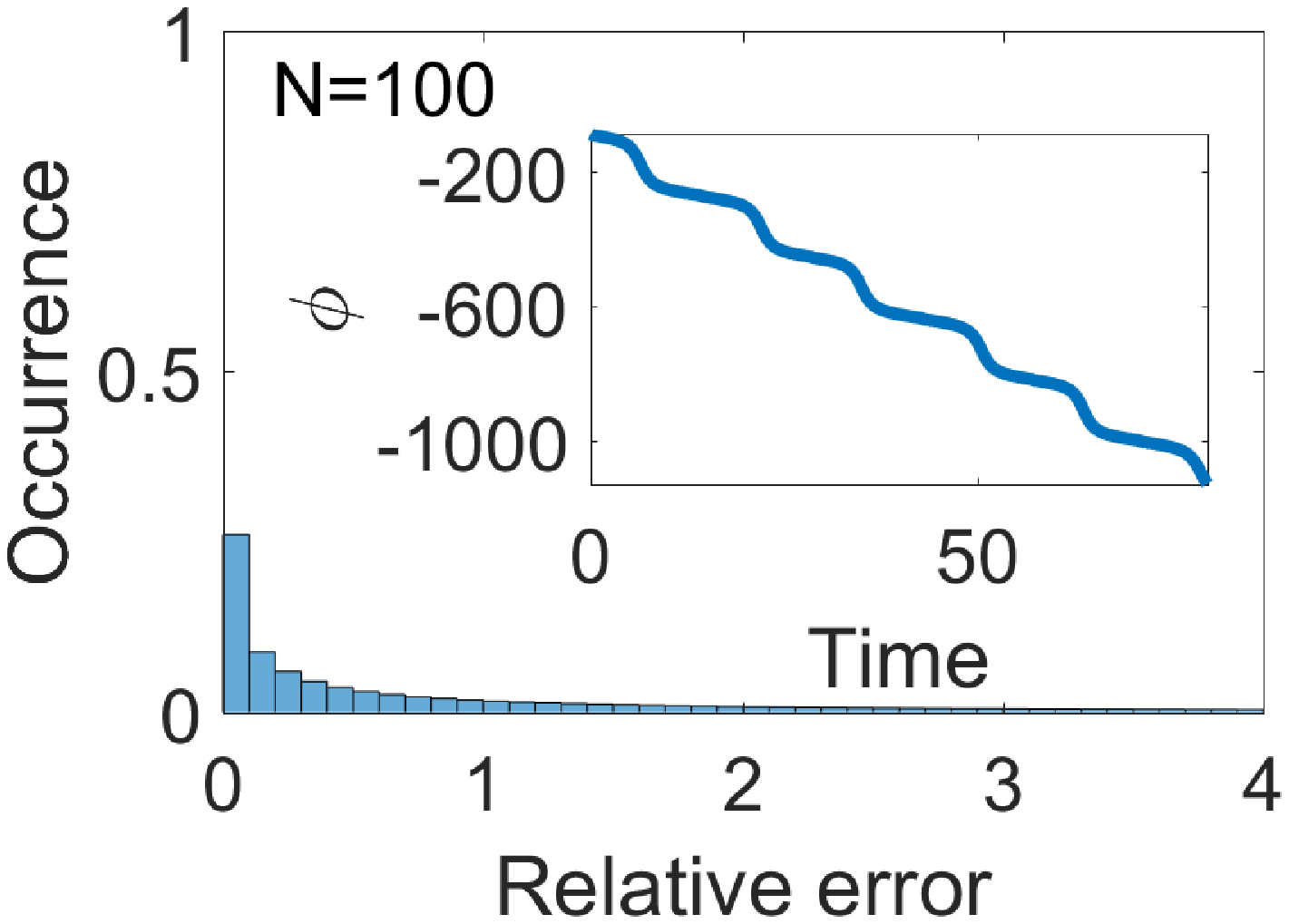}
	}
	\centerline{(a)}
	\subfigure
	{
	\includegraphics[width=2.1in] {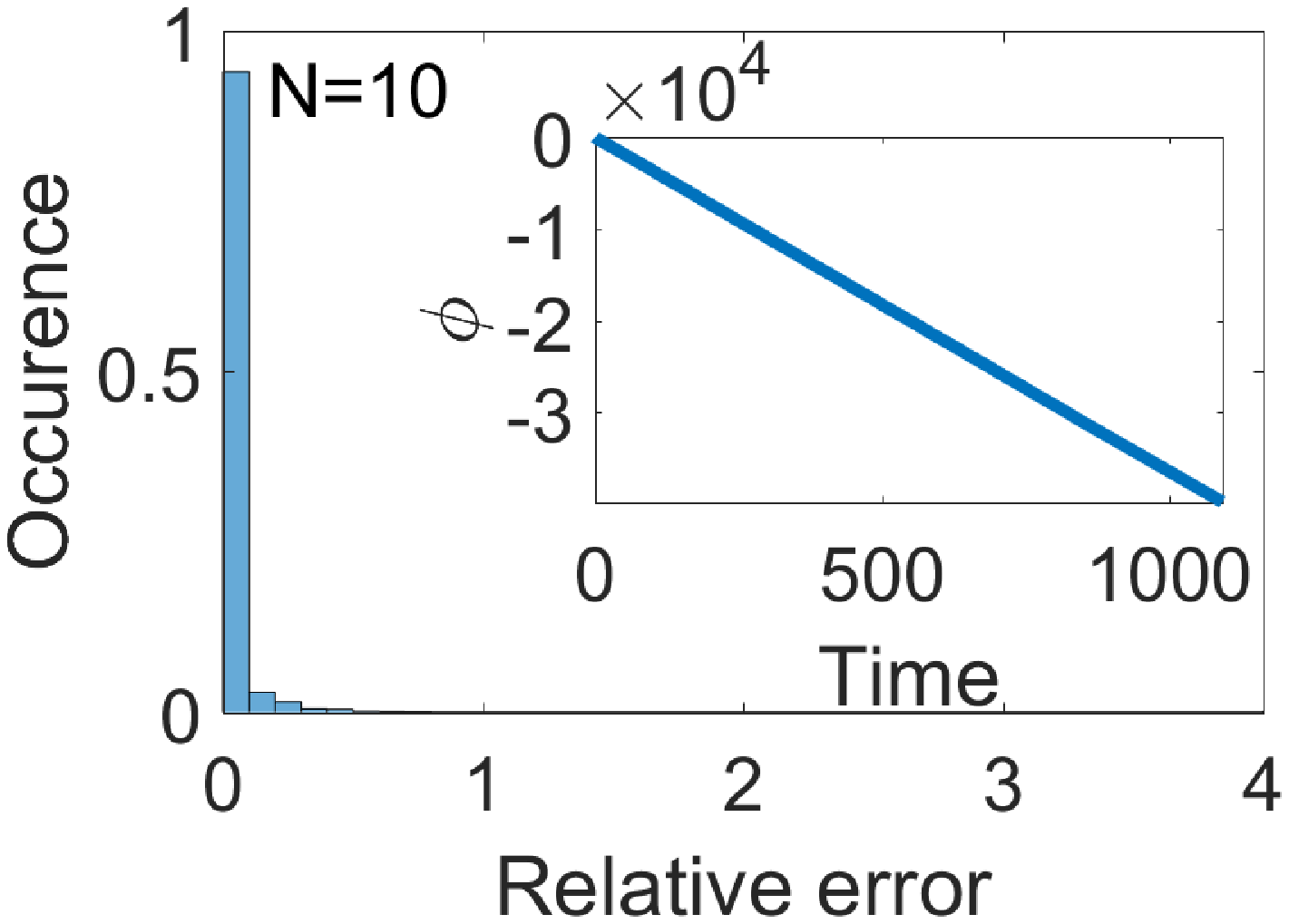}
	}~
	\subfigure
	{
	\includegraphics[width=2.1in] {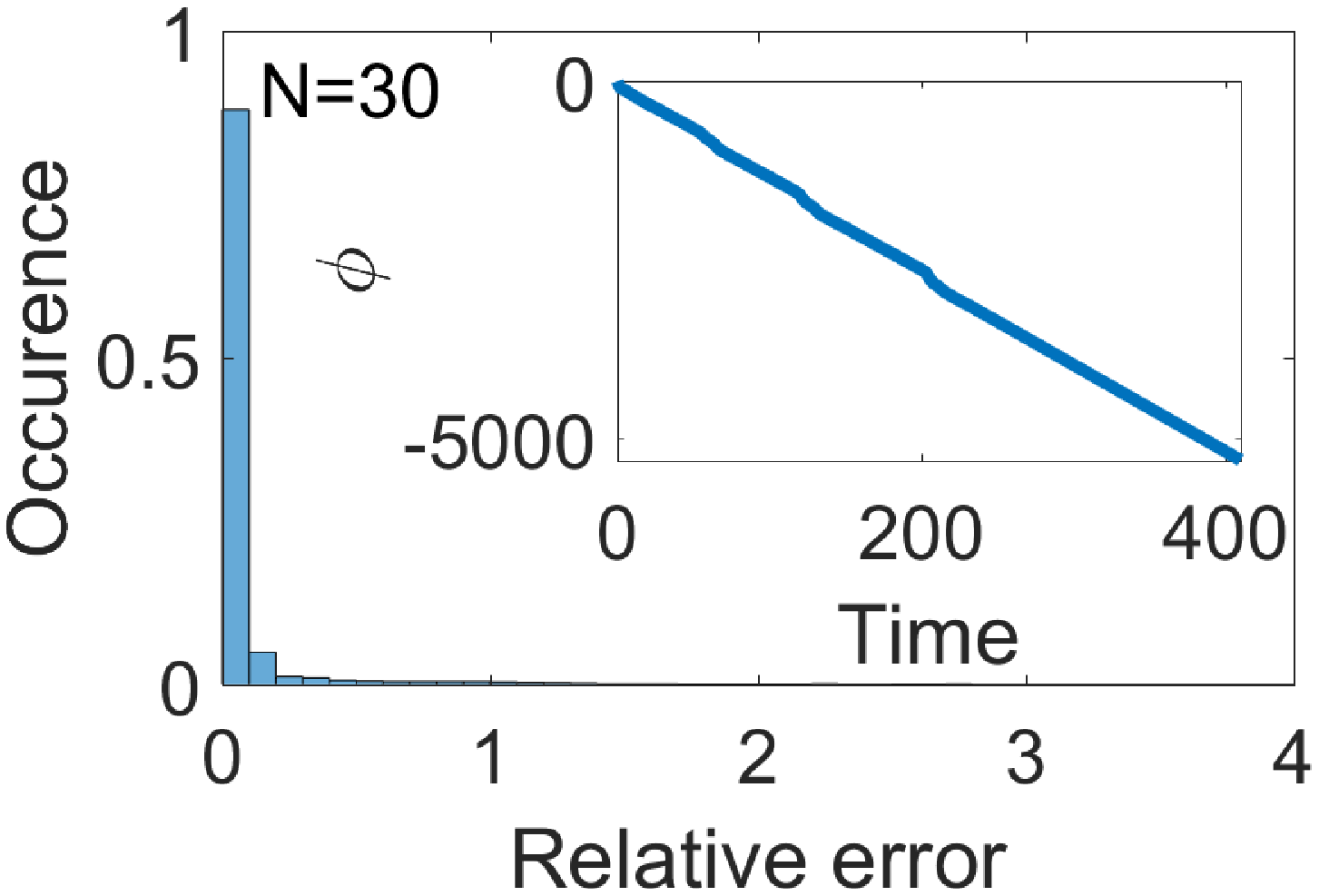}
	}~
	\subfigure
	{
	\includegraphics[width=2.1in] {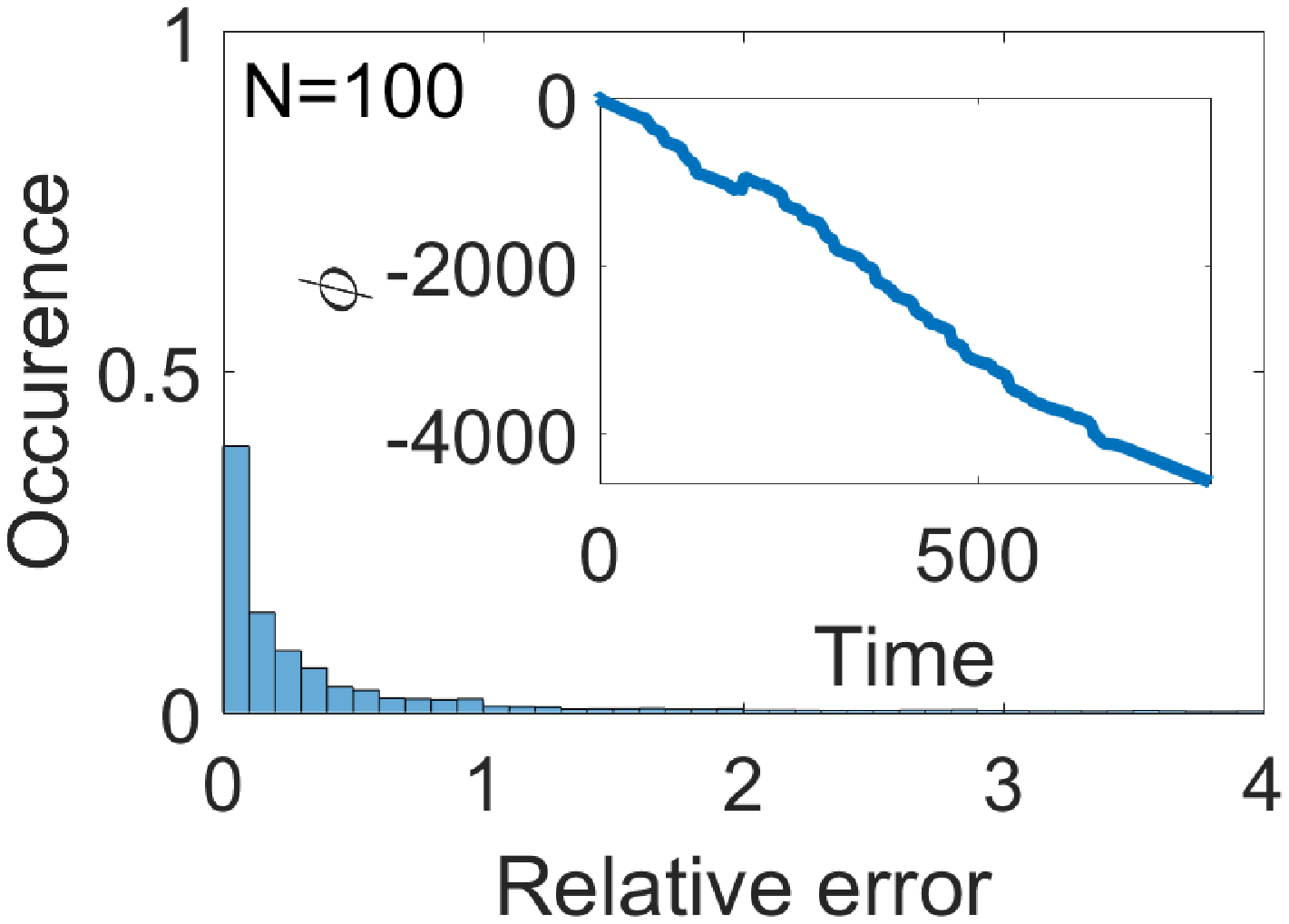}
	}
	\centerline{(b)}
\caption{Reduced order model may fail to hold in the DNA model of large lengths.\newline\fontsize{9}{9}\selectfont{ Figure shows the normalized frequency distribution of the relative error in the modal frequency for DNA systems of different lengths. The frequency is calculated by taking the derivative of $\phi$ using the first principle. The relative error in the x-axis is calculated as $|\frac{\dot{\phi}-\alpha_{1}}{\alpha_{1}}|$. The occurrence in the y-axis is used to denote the height of the bars. Here, the height of each bar represents the relative number of cases in which the relative error is within the bin specified by the width of the bar. The inset in the figure shows the predicted phase $\phi$ as a function of time. The analysis is done under a different perturbing condition (\textit{a}) In each case, the first mode is excited such that the average energy per unit pendulum, $\frac{E_{1}}{N}=10^{-4}$. For the above condition, the factor $\frac{ah}{2}\sqrt{\frac{E_{1}}{N}}=0.035$. The duration of the simulation is $5\alpha_{1}^{-1}$. (\textit{b}) The local perturbation given to each of the DNA systems is such that the DNA just undergoes a flip. Two pendula are perturbed towards the repulsive region such that the final position lies within the range $\numrange[range-phrase=-]{0.4}{0.9}$ rads. The duration of the simulation is restricted to the flipping time.}}
\end{figure*}

 To test this, we perform a modal analysis on a DNA system with three different lengths $N=\,$10, 30 and 100. First, we excite the first mode of the system in such a way that the average energy per pendulum is the same in each case. Second, we analyze how close the frequency of the first mode remains to its unperturbed value. We do this by applying the actual modal data to the reduced order model and predicting the phase, $\phi$, of the first mode using the following relation derived from\,(\ref{mode_sol}):

\begin{equation}
\phi(t)=-\frac{1}{\alpha_{1}}tan^{-1}\frac{\dot{\bar{\theta}}_{1}}{\bar{\theta}_{1}(t)},
\end{equation}

\noindent where $\dot{\bar{\theta}}_{1}$ and $\bar{\theta}_{1}(t)$are obtained from (\ref{modal_eq}). Next, we track the rate at which the phase, $\phi\left(t\right)$, changes to predict the perturbed modal frequency. The frequency distribution plots in Figure 4(a) show the deviation of the predicted modal frequency from its unperturbed value corresponding to DNA models of different lengths. The data shows that as $N$ increases, the relative error also increases. These deviations are observed to occur for a certain period when the slope of the modal phase becomes relatively steeper (please see the inset of Figure 4(a)). Further, we test the prediction made above for finite local perturbation as in Section $\ref{Flipping_dynamics_sec}$., where the perturbing condition is chosen such that the system simply flips (Figure 4(b)). The results are consistent with the trend in Figure 4(a), although the relative values are lower. Taken together, we infer that the reduced order model might not hold for large $N$, as the modal frequencies are more likely to deviate from their unperturbed value. These results suggest that for the large length models, resonant interaction between the modes may have to be considered to explain the flipping dynamics.

\section{DNA length can influence flipping behavior}\label{DNA_size_inf_flip_behavior}
The foregoing analysis indicates that the length of a DNA molecule can influence its flipping mechanism. To understand how the length might affect a DNA’s flipping behavior, we subject the DNA model of varying lengths to local perturbation and test their properties related to the flipping behavior. One such property is the energy threshold, which is the minimum energy required for flipping to take place. A possible implication of the DNA length on the energy threshold can be observed in the following example: Consider a case where all the pendula are perturbed such that they are equally pushed towards the repulsive region at the same time. Under such a condition, the energy threshold can be calculated to be~\cite{Toit}:

\begin{equation}\label{E_min}
E_{min}=N\,\epsilon\,\left(e^{-a\left(2h-x_{0}\right)}-1\right)^{2}=7.14\,N\,\times\,10^{-4},
\end{equation}
\noindent We note from the above expression that the energy threshold is proportionally related to the DNA length, N. Since the energy threshold inherently depends on the nature of the perturbation~\cite{Mezic}, the above condition may not be directly applicable to a local perturbation process. However, the condition may restrict the energy threshold for flipping to happen in such processes. To understand it, we select a local perturbation where a group of adjacent pendula, less than $N$, are targeted and are equally pushed towards the repulsive region. We call this process a `uniform local perturbation'. For example, let $m$ number of adjacent targeted pendula, less than $N$, be perturbed such that their angular positions are shifted equally to $\theta=0\degree$. Under this condition, the maximum perturbation energy, $E_{L}$, that can be transfered to the system is derived from $\left(\ref{Hamiltonian}\right)$ as:

\begin{equation}
\begin{split}
E_{L}&=E\left(\dot{\theta}_{1}\left(0\right)=0,\dot{\theta}_{2}\left(0\right)=0,\cdot\cdot,\theta_{1}(0)=0^{0},\theta_{2}(0)=0\degree,\cdot\cdot\theta_{m}\left(0\right)=0\degree,\theta_{m+1}\left(0\right)=\theta_{0},\cdot\cdot\right)\\&-E\left(\dot{\theta}_{1}\left(0\right)=0,\dot{\theta}_{2}\left(0\right)=0,\cdot\cdot,\theta_{1}(0)=\theta_{0},\theta_{2}(0)=\theta_{0},\cdot\cdot\right)=\theta_{0}^{2}+\epsilon\,m\,U\left(\theta=0\degree\right),
\end{split}
\end{equation}

\noindent As $E_{L}$ is independent of the DNA length, $N$, flipping may not happen if $E_{L}<E_{min}$. Considering this fact, we analyze the flipping dynamics of the DNA system by scaling the number of targeted pendula along with its length. In the present case, we choose 2$\%$ of the total number of pendulums, $N$. For instance, the perturbation provided to each pendulum is such that the final deviated position lies within the range $\numrange[range-phrase=-]{0.4}{0.6}$ rad. Such local perturbation can approximately model an enzyme's interaction with a DNA molecule, where it can actively perturb the bases while sliding along the strand at a sufficient speed~\cite{Toit}. 
\begin{figure*}[t!]
	\centering
	\subfigure[]
	{
	\includegraphics[scale=.33] {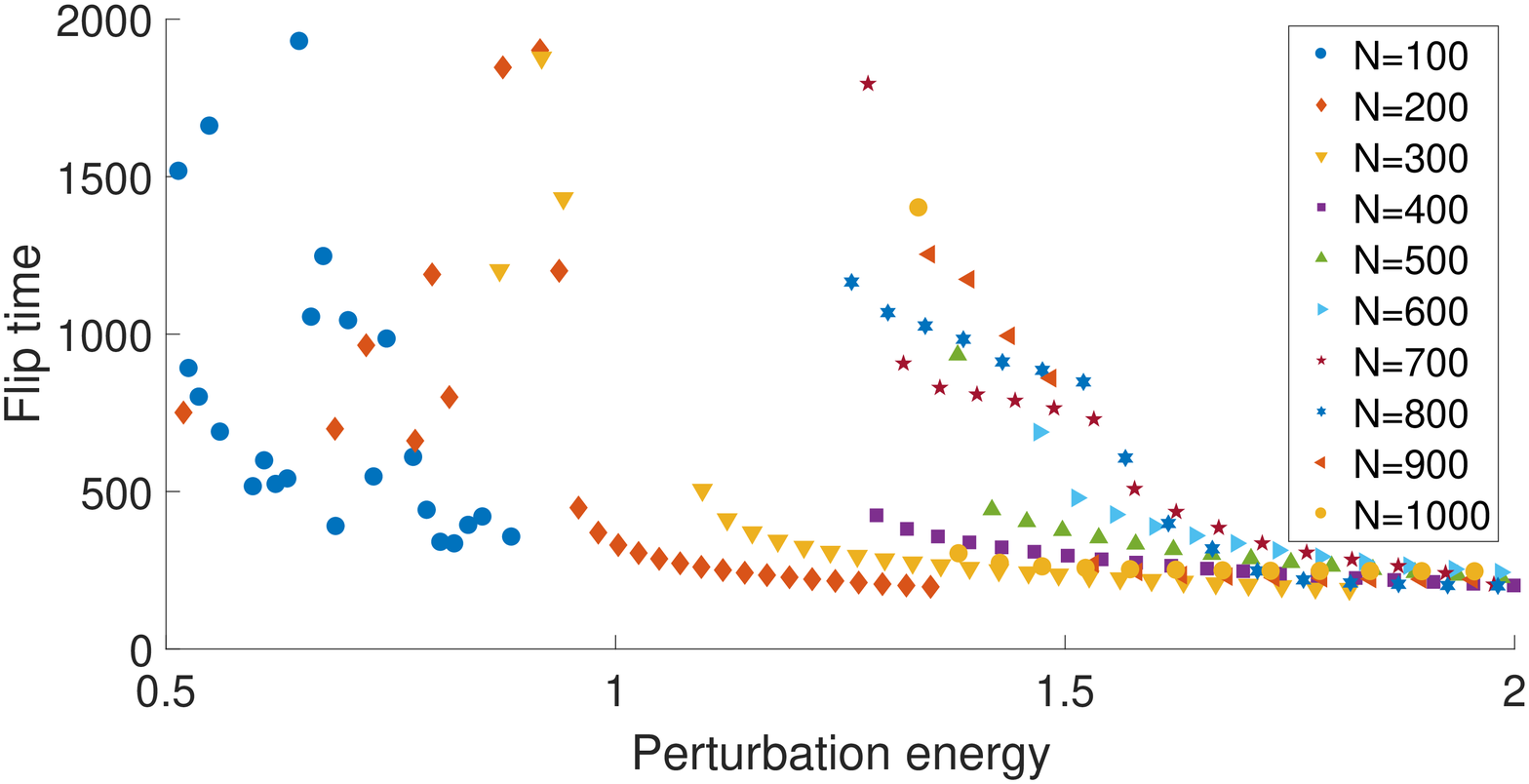}}~\centering
\subfigure[]
		{
\includegraphics[scale=.33]{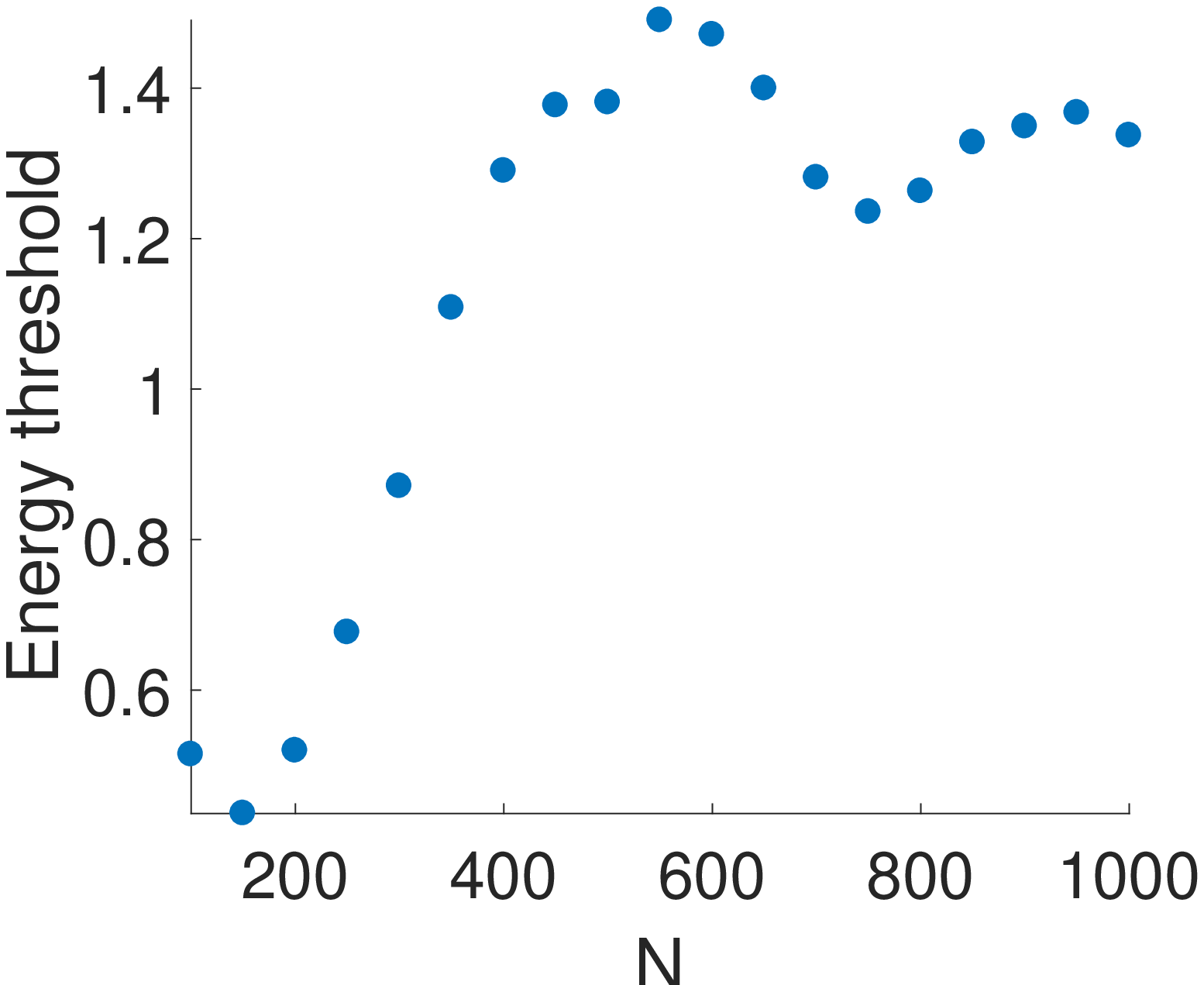}
		\label{fig:10percent}
		}
		\subfigure[]
		{
\includegraphics[scale=.45]{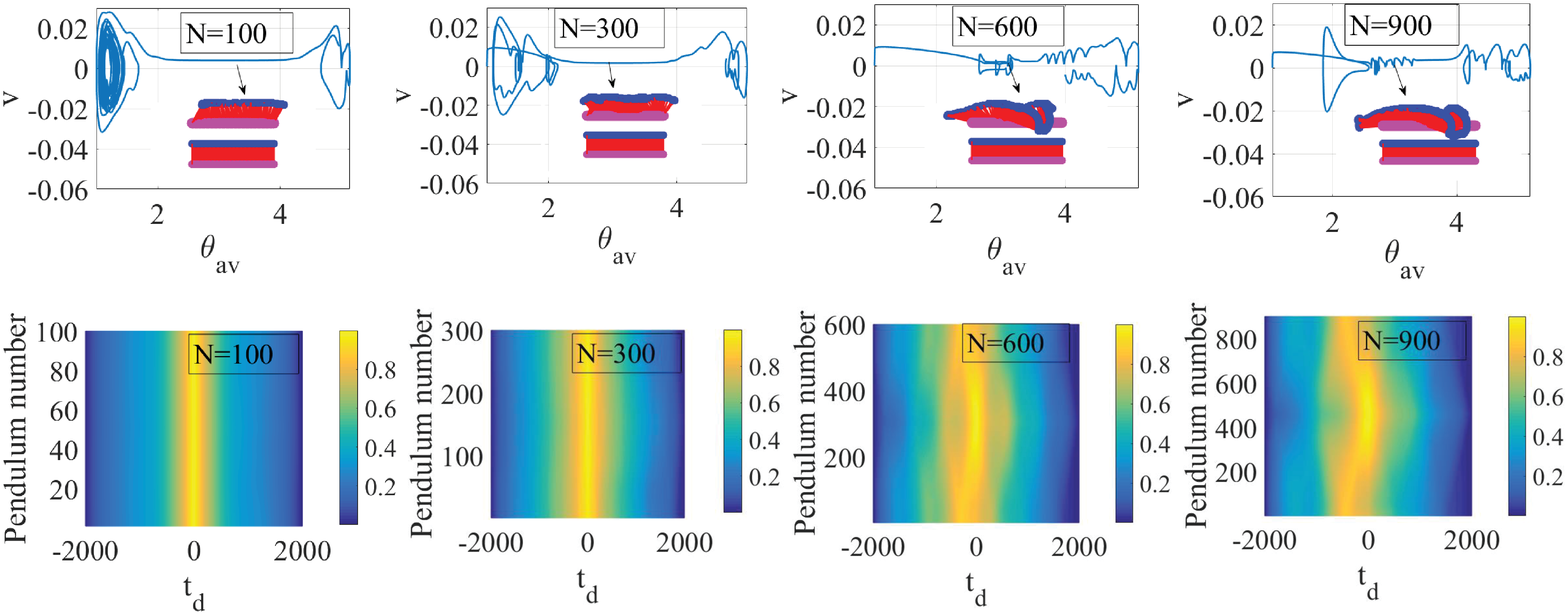}
		\label{fig:10percent}
		}		
	\caption{Flipping properties changes with increase in DNA length.\newline\fontsize{9}{9}\selectfont{(a) Scatter plot showing the time required to flip as a function of perturbation energy. For a DNA of a given length, 50 uniform local perturbations are given to $m=2\%$ of $N$ number of pendula. Each perturbation shifts the targeted pendula to a new position within a range $\numrange[range-phrase=-]{0.4}{0.6}$. For each case of perturbation, the duration of the simulation is fixed to 2000 units.(b) Scatter plot showing the energy threshold as a function of $N$. (c) Average phase portrait obtained for $N$= 100, 300, 600 and 900, when a flip is obtained at the threshold point. The inset in the average phase portrait plot shows a snapshot of the DNA model configuration at the flipping point, where the average co-ordinate crosses the $\theta_{av}=\pi$ mark at the first instance (indicated by the arrow). The cross-correlation coefficient $\rho$, corresponding to each $N$  is also plotted.}}
	\label{DNAmodel1}
\end{figure*}\noindent 

Figure 5 shows the flipping behavior of the DNA model of selected lengths (within a range, $N=\numrange[range-phrase=-]{100}{1000}$), subjected to uniform local perturbation. We start with studying the variation of the flip time with the perturbation energy for DNA models of different lengths, as shown in Figure 5(a). For a given length and within the perturbation energy window of $\numrange[range-phrase=-]{0.5}{2}$ units, it is observed that a large perturbation energy corresponds to a smaller flip time. Additionally, for a model of a given length, we find the existence of an energy threshold, below which the flip event is not observed. Also, we notice that these observations are in line with the pattern found in~\cite{Mezic,Toit}, where such flipping properties of a similar model subjected to structured perturbation were studied. We also find that, for the uniform local perturbation case, the energy threshold changes with $N$, as shown in Figure 5(b). Interestingly, it shows two different regimes where the DNA model may be operating: First, in the length range $N$ $\approx\numrange[range-phrase=-]{100}{600}$ where the energy threshold is observed to increase with $N$ for most of the range and second,in the length range $N$ $\approx\numrange[range-phrase=-]{600}{1000}$ where the threshold remains almost constant. Note that this is different from a result reported by a similar analysis in Mezic's work\cite{Mezic}, where a similar model was studied under half perturbation (number of targeted pendula being 50$\%$ of $N$) but for smaller length models. In his work, the flipping behavior was seen to be robust over the range of $N$ considered.

 Further to understand the flipping behavior at the threshold point, we projected the trajectory of the DNA system on to the average co-ordinates, corresponding to four different DNA lengths, as shown in Figure 5(c). For N = 100 and N = 300, we found the perturbed system initially remains in a `breathing state' for a certain period, before undergoing a flip to the other equilibrium point. Also, we computed the cross-correlation of the motion of the centre pendulum with the other pendula, finding that their motions are well correlated and approximately follow a rigid body dynamics similar to the dynamics obtained in the Section $\ref{Flipping_dynamics_sec}$ example. We also notice that during the transition period, the angular velocity is lower for $N=300$, as compared to $N=100$. A similar trend was seen in~\cite{Robinson}, where a low-resolution model of a DNA is considered and the motion of the bases is subjected to stochastic fluctuation. They found that the average time taken by the molecule to rotate by a fixed amount increases with the number of base pairs in the DNA molecule. In the second regime, however, we found a significant change in the flipping behavior. Unlike in case of the previous length models belonging to regime I, in regime II, the motion of the system during the flipping period deviates from the rigid body type motion. A possible evidence for this can be seen in the correlation plot shown in Figure 5(c), where the motion of pendula which are relatively far from the centre pendulum are less correlated. Although the distant pendula, such as the centre and the first pendulum are weakly correlated during simulation, a relatively good correlation exists between them when the motion of the centre pendula is observed after a certain lag. For instance, in case of $N=900$, the lag, $t_{d}\approx500$. This may indicate rotational deformation of the DNA model along its length. Further evidence of the deformation can be seen in the inset of Figure 5(c), which shows the DNA in a locally unwinded state~\cite{Yakushevich} at the flipping instant. Note that the flipping behavior in this regime is comparatively different from the behavior shown in similar models in earlier works~\cite{Mezic,Toit}, which show a rigid behavior as observed in the lower length models of regime I.
\begin{figure*}[t!]
	\centering
	\subfigure[]
	{
	\includegraphics[scale=.25] {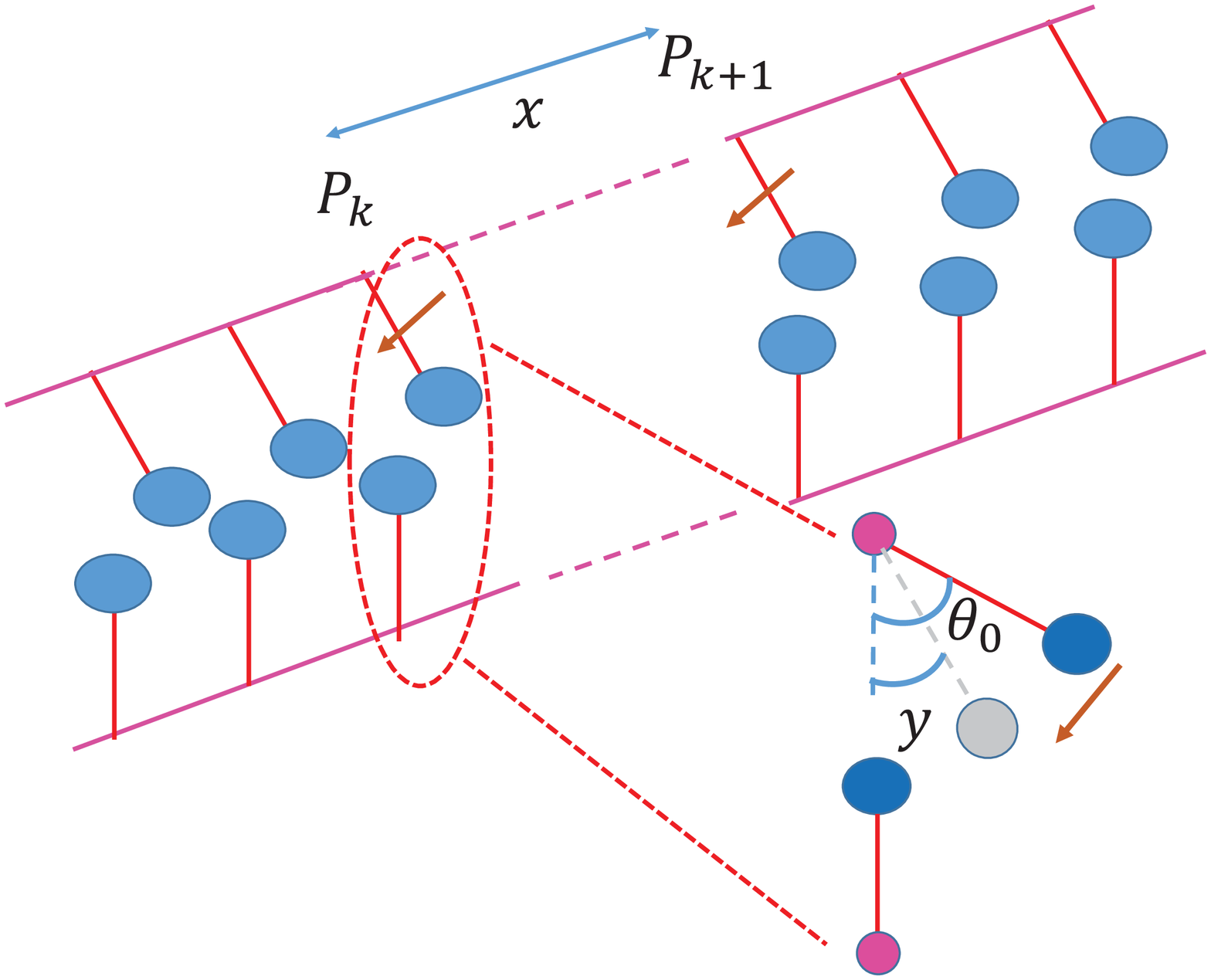}
	}~
\subfigure[]
		{
\includegraphics[scale=.28]{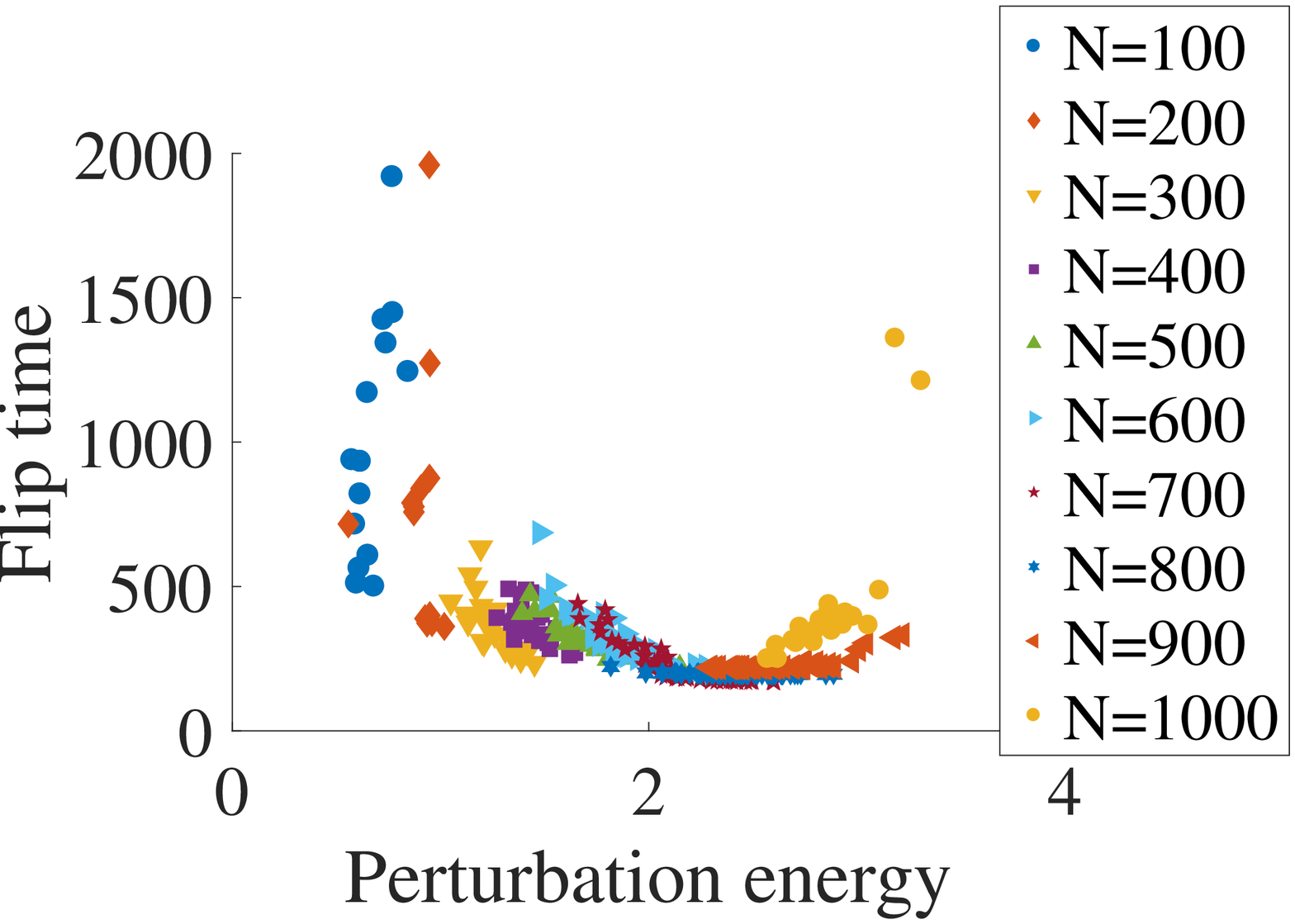}
		}
		
		\subfigure[]
		{
\includegraphics[scale=.45]{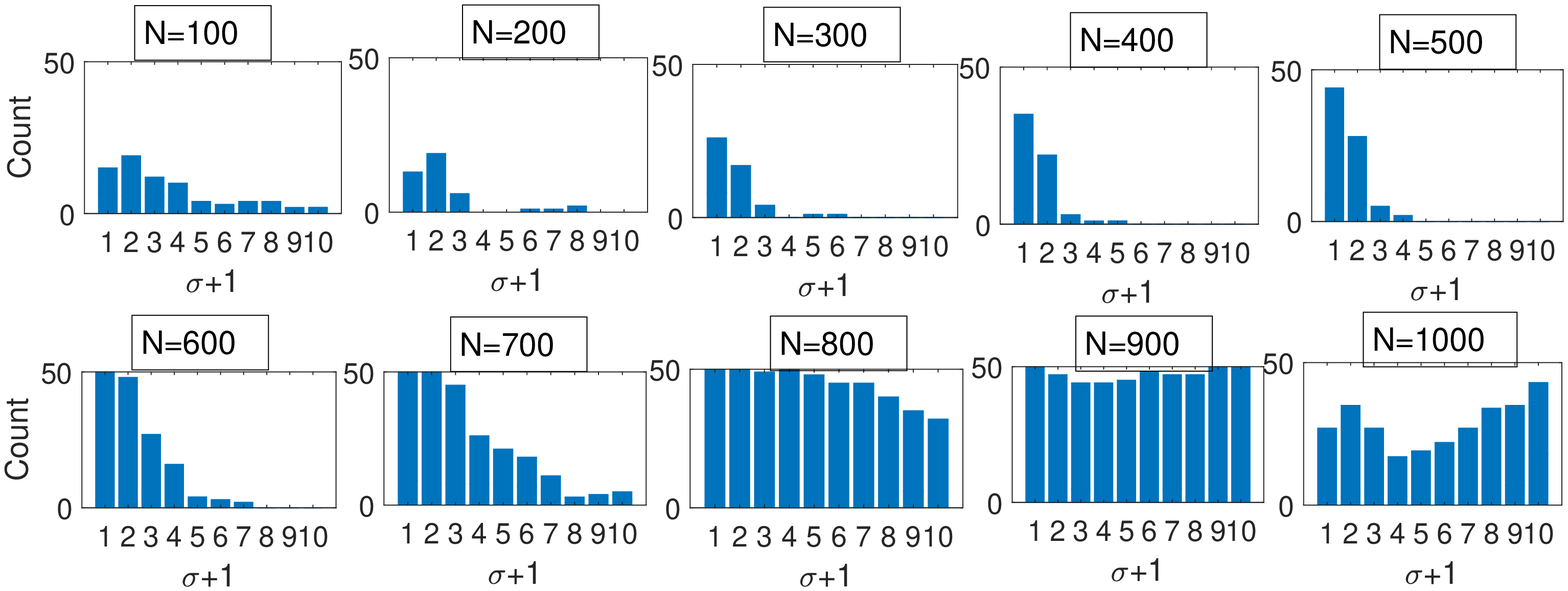}
		}		
	\caption{Flipping becomes more robust to random perturbation with the increase in DNA length. \newline\fontsize{9}{9}\selectfont{(a) Figure shows a schematic of a modified model of the local perturbation. For $m$ number of targeted pendula, the spacing between the $k^{th}$ and $k+1^{th}$ pendulum is $x-1$, where $x$ is an integer chosen randomly within a range 1 \textemdash$\,\sigma+1$ following a uniform probability distribution $f_{m}^{\sigma+1}\left(x\right)=\frac{1}{\sigma+1}$, where $\sigma$ is the maximum spacing  between the consecutive pendula selected. Each chosen pendulum is given a push towards the repulsive region where its new position is a random variable $y$ with a uniform probability distribution $f_{m}^{M}\left(y\right)=\frac{1}{M}$.Here,$M$ is the total number of samples that the angular range of $y$ is divided into. (b) Scatter plot showing the flipping time of the DNA system as a function of perturbation energy for zero maximum spacing ($\sigma=0$). In each case of $N$, 50 perturbations are given and for each perturbation, the number of pendula targeted is $m=2\%$ of $N$ and each targeted pendulum is deviated to a new angle within a range $\numrange[range-phrase=-]{0.4}{0.6}$ rad divided into $M=$50 segments. (c) Histogram plot showing the frequency distribution of the number of flips as a function of maximum spacing $\sigma$ for different DNA lengths.}}
	\label{DNAmodel1}
\end{figure*}

In the above computation, we studied the flipping behavior of the DNA model to uniform local perturbation where adjacent target pendula are chosen and identically perturbed. However, from a biological perspective, such a perturbation process can be inherently stochastic. For instance, in the case of an enzyme interacting with a DNA molecule, the process may model an enzyme taking a random step size during the translocation process~\cite{Patel} and perturbing the bases as it moves. A possible way of adding this in the DNA model can be through a random choice of the initial conditions. In~\cite{Mezic,Toit}, the flipping properties of a similar DNA model, with smaller lengths, have been tested by subjecting it to such random perturbations. Following these analyses, we next investigate how randomness in the local perturbation process may affect flipping in relatively larger length models. 

We start by introducing randomness in both the steps of the local perturbation process— in the selection of the target pendula and in the amount of deviation given to each of the selected pendula. We describe the method using the following rules:
\begin{itemize}
\item[1.]  Let $m$=2$\%$ of $N$ pendula be targeted for perturbation. If $P_{k}$ represents the position of the $k^{th}$ target pendulum, the position of the $k+1^{th}$ target pendulum is given as:
\begin{equation}
 P_{k+1}=P_{k}+x\qquad1\leq k\leq m-1\quad x\in[1,2,...\sigma+1],
\end{equation}
 where, $P_{1}=1$, $x$ is a random variable following a uniform discrete probability distribution, $f_{m}^{\sigma+1}\left(x\right)=\frac{1}{\sigma+1}$ and $x-1$ are the spacing between the two adjacent target pendula (see Fig 6(a)). A zero spacing between the pendula means that they are adjacent to each other. We further provide an approach to control the randomness in this process. In order to control the randomness, we first measure it using the Boltzmann-Shannon entropy~\cite{Kvalseth}, $H$. For a given sample length $\sigma+1$, the entropy is given as $H=-\sum_{i=1}^{\sigma+1}p_{i}log_{e}p_{i}=log_{e}\left(\sigma+1\right)$, where, $p_{i}=f_{m}^{\sigma+1}\left(x\right)=\frac{1}{\sigma+1}$. Note that as the randomness associated with this process is directly related to the sample size $\sigma+1$, we use the $\sigma$ parameter to control it. For instance, if $\sigma=0$, it corresponds to zero entropy or randomness in the base selection process. Also, note that as the average spacing between the targeted pendulum is $\frac{\sigma}{2}$, the randomness in the perturbation step can be directly related to how localised the perturbation is.
\item[2.] After selection, each targeted pendulum is pushed into the repulsive region where its new position is a random variable $y$ following a uniform discrete probability distribution $f_{m}^{M}\left(y\right)=\frac{1}{M}$, where $M=50$ is the total number of segments into which the angular range of $y$ is divided (see Figure 6(a)).
\end{itemize}

First, we study the case where the adjacent pendula are targeted and pushed randomly within the angular range $\numrange[range-phrase=-]{0.4}{0.6}$ rad. In this case, for different length models, we compute the variation of flip time with the perturbation energy (Figure 6(b)). We find that the lower length models $\left(
N\approx\numrange[range-phrase=-]{100}{600}\right)$ show similar flipping properties when compared to the case where they are subjected to uniform local perturbation (Figure 5(a)). However, when the properties of the larger length models are compared, the pattern is observed to be different, with a lower frequency of flip events within the energy range $\numrange[range-phrase=-]{0.5}{2}$ units. Second, we introduce randomness in the base selection process and investigate how it may influence the flipping behavior. For each length of the DNA model and a given entropy in the base selection process, we provide 50 random perturbations to the selected bases using rule 2 and count the occurrence of a flip event. The result obtained in Figure 6(c) suggests that the system with a lower length $\approx\numrange[range-phrase=-]{100}{600}$ is relatively immune to conformational change when subjected to perturbation with large randomness. This is different from the case where the randomness in the perturbation process is relatively low, which shows more flipping events. This could be because when randomness is reduced, the perturbation tends to be more local, which increases the chances of flipping in the model~\cite{Toit}. However, in the case for other lengths, the trend does not hold and the flip event becomes relatively sensitive to random perturbation. Note that although the observation for the lower length models is in line with the trend seen in~\cite{Mezic,Toit,Eisenhower}, it is different for the larger length models. This result indicates that when perturbed randomly, $N$ may play a vital role in determining the system's tendency to remain in a stable conformation.

\section{Discussion} Developing coarse models of flipping in a DNA molecule is important in understanding how structural features of the molecule could influence its dynamics. In this paper, we investigated whether and how the DNA length affects the applicability of a previously developed reduced order model and the flipping properties of the full model under local perturbation. Using the modified Lindstedt-Poincare method, we investigated the perturbed frequency of the modes close to the equilibrium point for single mode perturbation. Our findings suggest that the approximations on which the reduced order model is based may not hold for DNA models with sufficiently large lengths as the modal frequency is observed to deviate significantly from their unperturbed value. Further, in order to understand how an increase in the DNA length may affect flipping behavior, we numerically simulated the full model with comparatively large lengths by subjecting it to uniform local perturbation. For the given parameter and input properties, we found that for most of the range, N$\approx\numrange[range-phrase=-]{100}{600}$, the threshold energy required to flip increases with N, whereas after N$\approx$600, the energy remains almost constant. Finally, we found that with the increase in the DNA length, the propensity of the system to remain in a stable conformation against random perturbation decreases.

It is interesting to note that the length of a DNA molecule can influence its effective rigidity. For lower lengths, the DNA model is essentially seen to follow rigid body dynamics, whereas it behaves as a flexible body for larger lengths,where the DNA backbone is seen to undergo rotational deformation along its length. This change in behavior is similar to the change observed in the bending property of a semi-flexible polymer, modeled as a worm-like chain~\cite{Marantan}, when its length crosses a characteristic bending length scale called the persistence length~\cite{Calladine}. For instance, for lengths shorter than the persistence length, the molecule shows high bending rigidity, whereas for longer lengths it behaves as a flexible body which can be bent easily. A possible reason for the change in the bending behavior could be the strong influence of the low frequency modes on the overall motion of the DNA system~\cite{Matsumoto}. In fact, the strong deformation observed in our model could be the result of a considerable amount of energy being funneled into the lower modes as compared to the higher modes. According to our study, this may occur because of the large perturbation the frequency of the lower modes undergo, which makes it more likely to be commensurate with the frequency of the higher modes, resulting in resonance and energy exchange between them.  

Note, we have chosen a conservative chain of coupled oscillators for modelling DNA conformational change because similar models have been previously used as a first approximation towards building a more accurate theoretical model of DNA internal dynamics~\cite{Yakushevich2}. Moreover, the goal of this paper is to study the internal dynamics of the DNA molecule, which is largely conservative in nature. Interaction of a DNA molecule with its environment and other cell constituents may be better modelled as a dissipative process, but that is outside the scope of our work.

An essential feature of the conservative model observed in our paper is that for a given nature of perturbation, there exists a threshold energy below which no conformational change happens. Similar evidence of such property can be found in phenomenon such as DNA thermal denaturation or melting, which involves the separation of two strands of a DNA molecule by heating. Specifically, observations using spectroscopic methods~\cite{Owczarzy} have shown the existence of a threshold temperature (or melting temperature) beyond which a DNA molecule undergoes a structural transition  from a double-stranded form to a single-stranded form.

Another useful feature of the DNA model is the influence of its length on its mechanical stability. A particular instance of this length-dependent feature can be observed in the pattern of the curve in Figure 5(b) that depicts an initial increase in energy threshold with chain length that further approximately saturates to a constant level. The above result supports evidence of a similar pattern observed in thermal denaturation studies done under constant physiological conditions~\cite{Manyanga,Manyanga2}, where for shorter DNA lengths, the melting temperature initially rises with the number of nucleotides present in the molecule following which the melting temperature saturates to a fixed level for longer lengths. Evidence of such length dependent feature can also be found in~\cite{Singh}, where the mechanical stability of a heterogeneous ds-DNA molecule was investigated using a coarse-grained model. Similar to our result, it shows the existence of a minimum length beyond which the length of a DNA molecule has minimal contribution to its mechanical stability.

Although the model considered in our work shows similar DNA mechanical properties as observed in experimental conditions, the results obtained using the model may not be directly applicable. This non-applicability is because in our case, energy is provided to the system via perturbations that are spatially localized along its length, whereas, in experimental conditions, the perturbations that act on the molecule are not spatially restricted. A possible area where our work would be useful is in understanding the mechanical stability of a DNA molecule subjected to an external force acting locally on specific sites. These forces may be provided using single-molecule experimental tools~\cite{Ritort}. However, to our knowledge, applications of these tools are limited to exerting force at the ends of the DNA molecule.

 Apart from the length of the molecule, there are other physical factors, such as the stiffness of the bonds present in the DNA molecule, which can influence the DNA internal dynamics. This is evident in the perturbation term of the modal frequency expression~$\left(\ref{pert_freq}\right)$ obtained for the single mode perturbation case. For instance, in the present case, two factors on which the stiffness depends are \textemdash$\,$ $\epsilon$, which indirectly depends on the strength of the torsional interaction in the molecule's backbone, and the decay coefficient $a$, which is directly related to the stiffness of the hydrogen bonds between the base pairs. Our future work would be to investigate whether the suitable value of these parameters could be obtained such that the reduced order model would hold when the length of the full model is large.  

In conclusion, our work highlights the importance of the length of a DNA model in controlling its flipping dynamics.This paper may also help us understand how an enzyme can use this property of the molecule to efficiently trigger its unwinding.

\section{Acknowledgements}
S.B. would like to thank the Visvesvaraya Ph.D. Scheme for Electronics $\&$ IT, Ministry of Electronics and Information Technology (MeitY), Government of India, for financial support (File No. IITD/IRD/MI01233).

\section{Appendix}
\noindent Consider in the DNA model all the pendula are initially resting at the equilibrium angular position, $\theta_{0}$. Next, we excite its $w^{th}$ nonzero mode through the following set of initial conditions,
\begin{equation}
\label{modes_IC}
\begin{split}
\left(\bar{\theta}_{0,w'}(0),\dot{\bar{\theta}}_{0,w'}(0)\right)&=\left(0,\sqrt{2E_{w}}\right)\quad w'=w\,and\,w'\neq0,\\&=\left(0,0\right)\quad\;\qquad\,\,\,w'\neq w\,and\,w'\neq0,\\&=\left(\sqrt{N}\theta_{0},0\right)\;\qquad\quad\enskip w'=0,
\end{split}
\end{equation}
\noindent Note, here $E_{w}$ is the energy imparted to the $w^{th}$ mode at time $t=0$. Under the aforesaid condition, the $\epsilon^{0}$ order solutions of the modes can be written as,
\begin{equation}
\begin{split}\epsilon^{0}\, & :\qquad\qquad\,\bar{\theta}_{0,w'}=A_{w'}\,sin\left(\Omega_{w'}t\right),\quad\,w'=w\,\mathrm{and}\,w\neq0\\
 & \,\qquad\qquad\,\,\,\quad\quad=0,\qquad\qquad\quad\quad\quad w'\neq w\,\mathrm{and}\,w'\neq0\\
 & \,\qquad\qquad\,\,\,\quad\quad=\sqrt{N}\theta_{0},\qquad\quad\qquad\;\,\,\quad w'=0
\end{split}
\end{equation}
\noindent where, $A_{w}=\sqrt{2\,E_{w}}$. Next, using $\left(\ref{mode_eq_order}\right)$ and $\left(\ref{modes_IC}\right)$ we obtain the $\epsilon$ order dynamics of the $w^{th}$ nonzero mode given as
\begin{equation}
\label{theta_w_1_dynamics}
\epsilon:\qquad\ddot{\bar{\theta}}_{1,w}+\Omega_{w}^{2}\,\bar{\theta}_{1,w}=-\Omega_{w,1}^{2}\,A_{w}\,sin\left(\Omega_{w}t\right)-\sum_{n=1}^{N}T_{nw}G\left(\theta_{0}+T_{nw}A_{w}\,sin\left(\Omega_{w}t\right)\right),
\end{equation}
\noindent On expanding function $G\left(\theta\right)$ about $\theta_{0}$, in $\left(\ref{theta_w_1_dynamics}\right)$ we get, 
\begin{equation}
\label{theta_w_1_expand}
\begin{split}
\ddot{\bar{\theta}}_{1,w}+\Omega_{w}^{2}\,\bar{\theta}_{1,w}&=-\Omega_{w,1}^{2}\,A_{w}\,sin\left(\Omega_{w}t\right)-G\left(\theta_{0}\right)\sum_{n=1}^{N}T_{nw}-\partial_{\theta}G\left(\theta_{0}\right)A_{w}\,sin\left(\Omega_{w}t\right)\sum_{n=1}^{N}T_{nw}^{2}-,\\&\cdot\cdot\cdot\frac{1}{\left(k-1\right)!}\partial_{\theta}^{k-1}G\left(\theta_{0}\right)A_{w}^{k-1}\,sin^{k-1}\left(\Omega_{w}t\right)\sum_{n=1}^{N}T_{nw}^{k}-\cdot\cdot\cdot,
\end{split}
\end{equation}
\noindent Note here, $\partial_{\theta}^{k-1}G\left(\theta_{0}\right)$ represent $\partial_{\theta}^{k-1}G\left(\theta\right)$ evaluated at $\theta_{0}$. The second term on the right hand side vanishes as $G\left(\theta_{0}\right)=0$.
\noindent Next, we evaluate $\sum_{n}T_{nw}^{k}$ in $\left(\ref{theta_w_1_expand}\right)$. For $w<N/2$, each element of the $T$ matrix can be written in the form as $T_{nw}=\sqrt{\frac{2}{N}}cos\left(\frac{2\pi nw}{N}\right)$, then,
\begin{equation}
\sum_{n=1}^{N}T_{nw}^{k}=\left(\frac{2}{N}\right)^{k/2}\sum_{n=1}^{N}cos^{k}\left(\frac{2\pi nw}{N}\right),
\end{equation}
Let $\frac{2\pi nw}{N}=\phi$, then for even $k$, we have,

\begin{equation}
\label{sum_Tnw^k}
\begin{split}
\sum_{n=1}^{N}T_{nw}^{k}&=\left(\frac{2}{N}\right)^{k/2}\sum_{n=1}^{N}\frac{1}{2^{k}}\left[C_{\frac{k}{2}}^{k}+2\sum_{m=0}^{\frac{k}{2}-1}C_{m}^{k}cos\left(2m-k\right)\phi\right]\\&=\frac{1}{2^{\frac{k}{2}}}\frac{1}{N^{\frac{k}{2}}}\left[N\,C_{\frac{k}{2}}^{k}+2\sum_{m=0}^{\frac{k}{2}-1}C_{m}^{k}\sum_{n=1}^{N}cos\left(2m-k\right)\phi\right],
\end{split}
\end{equation}
Let $2m-k=z$, and replacing $\phi=\frac{2\pi nw}{N}$ above, we simplify the summation term over $n$ within the summation term over m
\begin{equation}
\begin{split}
\sum_{n=1}^{N}cos\left(z\phi\right)&=\sum_{n=1}^{N}cos\left(n\phi'\right)\qquad\phi'=\frac{2\pi zw}{N}\\&=Re\left[\sum_{n=1}^{N}e^{in\phi'}\right]\\&=Re\left[e^{i\phi'}\left(\frac{e^{iN\phi'}-1}{e^{i\phi'}-1}\right)\right]\\&=Re\left[e^{\frac{i\left(N+1\right)\phi'}{2}}\left(\frac{sin\left(\frac{N\phi'}{2}\right)}{sin\left(\frac{\phi'}{2}\right)}\right)\right]\\&=0,\qquad \mathrm{as}\,\,N\phi'=2\pi zw
\end{split}
\end{equation}
Using the above relation in $\left(\ref{sum_Tnw^k}\right)$, we obtain,
\begin{equation}
\sum_{n=1}^{N}T_{nw}^{k}=\frac{1}{2^{\frac{k}{2}}}\frac{1}{N^{\frac{k}{2}-1}}C_{\frac{k}{2}}^{k},
\end{equation}
For odd $k$, following the same procedure as above, we get,
\begin{equation}
\begin{split}
\sum_{n=1}^{N}T_{nw}^{k}&=\sum_{n=1}^{N}\left(\frac{2}{N}\right)^{k/2}\frac{1}{2^{k}}\left[2\sum_{m=0}^{\frac{k-1}{2}}C_{m}^{k}cos\left(2m-k\right)\phi\right]\\&=0,
\end{split}
\end{equation}
From above result the terms in $\left(\ref{theta_w_1_expand}\right)$ with odd $k$ vanishes. For even $k$, to calculate the coefficient of the secular term, we first extract the coefficient of $sin\left(\Omega_{w}t\right)$ from $sin^{k-1}\left(\Omega_{w}t\right)$.
Let $k-1=k'$ and $\Omega_{w}t=\Phi$, then,
\begin{equation}
\begin{split}
sin^{k-1}\left(\Omega_{w}t\right)&=sin^{k'}\left(\Phi\right)=\left(\frac{e^{i\Phi}-e^{-i\Phi}}{2i}\right)^{k'}\\&=\frac{1}{\left(2i\right)^{k'}}\sum_{m=0}^{k'}\left(-1\right)^{k'-m}C_{m}^{k'}e^{im\Phi}e^{-i\left(k'-m\right)\Phi}\\&=\frac{1}{\left(2i\right)^{k'}}\sum_{m=0}^{k'}\left(-1\right)^{k'-m}C_{m}^{k'}e^{i\left(2m-k'\right)m\Phi},
\end{split}
\end{equation}
The secular term can be obtained from the above expansion by combining the terms corresponding to $m=\frac{k'-1}{2}$ and $m=\frac{k'+1}{2}$ ,
\begin{equation}
\label{sin^kphi}
\begin{split}
sin^{k'}\left(\Phi\right)&=\frac{1}{\left(2i\right)^{k'}}\left[\cdot\cdot\cdot+C_{\frac{k'+1}{2}}^{k'}\left(-1\right)^{\frac{k'-1}{2}}e^{i\Phi}+C_{\frac{k'-1}{2}}^{k'}\left(-1\right)^{\frac{k'+1}{2}}e^{-i\Phi}+\cdot\cdot\cdot\right]\\&=\cdot\cdot\cdot+\frac{1}{\left(2i\right)^{k'-1}}C_{\frac{k'-1}{2}}^{k'}\left(-1\right)^{\frac{k'-1}{2}}\left[\frac{e^{i\Phi}-e^{-i\Phi}}{2i}\right]+\cdot\cdot\cdot\\&=\cdot\cdot\cdot+\frac{1}{\left(2i\right)^{k'-1}}C_{\frac{k'-1}{2}}^{k'}\left(-1\right)^{\frac{k'-1}{2}}sin\Phi+\cdot\cdot\cdot\\&=\cdot\cdot\cdot+\frac{1}{2^{k'-1}}C_{\frac{k'-1}{2}}^{k'}sin\Phi+\cdot\cdot\cdot,
\end{split}
\end{equation}
Combining the results of $\left(\ref{theta_w_1_expand}\right)$ and $\left(\ref{sin^kphi}\right)$ we get the coefficient of secular term which is equated to zero,
\begin{equation}
-\Omega_{w,1}^{2}\,A_{w}-\sum_{k=even,k>1}\frac{1}{\left(k-1\right)!}\partial_{\theta}^{k-1}G\left(\theta_{0}\right)A_{w}^{k-1}\frac{1}{2^{k-2}}C_{\frac{k-2}{2}}^{k-1}\frac{1}{2^{\frac{k}{2}}}\frac{1}{N^{\frac{k}{2}-1}}C_{\frac{k}{2}}^{k}=0,
\end{equation}
On simplifying the second term,
\begin{equation}
\begin{split}
\sum_{k=even,k>1}\frac{1}{\left(k-1\right)!}\partial_{\theta}^{k-1}G\left(\theta_{0}\right)A_{w}^{k-1}\frac{1}{2^{k-2}}C_{\frac{k-2}{2}}^{k-1}\frac{1}{2^{\frac{k}{2}}}\frac{1}{N^{\frac{k}{2}-1}}C_{\frac{k}{2}}^{k}=&\\\sum_{k=even,k>1}\partial_{\theta}^{k-1}G\left(\theta_{0}\right)A_{w}^{k-1}\frac{1}{2^{k-2}}\frac{1.3.5\cdot\cdot\cdot k-1}{\left(\frac{k}{2}!\right)^{2}\left(\frac{k}{2}-1\right)!}&\frac{1}{N^{\frac{k}{2}-1}}
\end{split},
\end{equation}
We obtain the perturbed frequency as,
\begin{equation}
\label{perturbed_freq_expansion_appendix}
\begin{split}
\Omega_{w}^{2}&=\alpha_{w}^{2}+\epsilon\,\partial_{\theta}^{2}U\left(\theta_{0}\right)+\epsilon\,\partial_{\theta}^{4}U\left(\theta_{0}\right)\,A_{w}^{2}\,\frac{3}{16}\,\frac{1}{N}+\cdot\cdot\cdot\\&+\epsilon\,\partial_{\theta}^{k}U\left(\theta_{0}\right)A_{w}^{k-2}\frac{1}{2^{k-2}}\frac{1.3.5\cdot\cdot\cdot k-1}{\left(\frac{k}{2}!\right)^{2}\left(\frac{k}{2}-1\right)!}\frac{1}{N^{\frac{k}{2}-1}}+\cdot\cdot\cdot,where\,k=2,4,6\cdot\cdot
\end{split},
\end{equation}
Next we find the order of the function $\partial_{\theta}^{k}U\left(\theta_{0}\right)$, we use the Fa\`a di Bruno's formula~\cite{Faadvi}, through which we write the $k^{th}$ derivative of $U\left(\theta\right)$ evaluated at $\theta=\theta_{0}$, in terms of derivative of $U$ with respect to a new variable, $z=\bar{h}\left(1-cos\theta\right)$. It is given as,
\begin{equation}
\label{Faadvi_expansion}
\partial_{\theta}^{k}U\left(\theta_{0}\right)=\sum_{m_{1},m_{2},\cdot\cdot m_{k}}\frac{k!}{m_{1}!m_{2}!\cdot\cdot m_{k}!}\partial_{z}^{K}U\left(z\left(\theta_{0}\right)\right)\,\left(\partial_{\theta}z\left(\theta_{0}\right)\right)^{m_{1}}\left(\frac{\partial_{\theta}^{2}z\left(\theta_{0}\right)}{2!}\right)^{m_{2}}\cdot\cdot\left(\frac{\partial_{\theta}^{k}z\left(\theta_{0}\right)}{k!}\right)^{m_{k}}
\end{equation}
where, $m_{1}$,$m_{2}$,$\cdot\cdot m_{k}$ are non-negative integers and $K=m_{1}+m_{2}+\cdot\cdot\cdot+m_{k}$, where the sum obeys a partition law given as,
\begin{equation}
m_{1}+2\,m_{2}+\cdot\cdot+k\,m_{k}=k,
\end{equation}
The $K^{th}$ derivative of $U$ with respect to $z$ inside the summation of $\left(\ref{Faadvi_expansion}\right)$ is given as,
\begin{equation}
\label{U_partial_z}
\partial_{z}^{K}U=2\left(\left(-1\right)^{K-1}e^{-\left(z-\bar{x}_{0}\right)}+\left(-1\right)^{K}2^{K-1}e^{-2\left(z-\bar{x}_{0}\right)}\right),
\end{equation}
As per the properties of asymptotic notations~\cite{Cormen}, the following can be expressed as, $\partial_{\theta}z\left(\theta_{0}\right)=\bar{h}sin\theta_{0}=\mathcal{O}\left(\bar{h}\right),\:\partial_{\theta}^{2}z\left(\theta_{0}\right)=\bar{h}cos\theta_{0}=\mathcal{O}\left(\bar{h}\right)\cdot\cdot\cdot\partial_{\theta}^{k}z\left(\theta_{0}\right)=\mathcal{O}\left(\bar{h}\right)$, then,
\begin{equation}
\begin{split}
\left(\partial_{\theta}z\right)^{m_{1}}\left(\frac{\partial_{\theta}^{2}z}{2!}\right)^{m_{2}}\cdot\cdot\left(\frac{\partial_{\theta}^{k}z}{k!}\right)^{m_{k}}&=\mathcal{O}\left(\bar{h}^{m_{1}}\right)\mathcal{O}\left(\bar{h}^{m_{2}}\right)\cdot\cdot\mathcal{O}\left(\bar{h}^{m_{k}}\right)=\mathcal{O}\left(\bar{h}^{m_{1}+m_{2}+\cdot\cdot\cdot+m_{k}}\right),\\&\qquad\mathrm{\left[If\,f_{1}=\mathcal{O}\left(g_{1}\right)\,and\,f_{2}=\mathcal{O}\left(g_{2}\right),\,then,\,f_{1}f_{2}=\mathcal{O}\left(g_{1}g_{2}\right)\right]}
\end{split},
\end{equation}
As, $m_{1}+m_{2}+\cdot\cdot\cdot+m_{k}\leq m_{1}+2\,m_{2}+\cdot\cdot+k\,m_{k}=k$, then of all the combination of $m_{1},m_{2},\cdot\cdot m_{k}$, the combination which gives, $m_{1}+m_{2}+\cdot\cdot\cdot+m_{k}=k$, determines the order of the above expression. Also, the value of $m_{1}, m_{2},\cdot\cdot m_{k}$, which would satisfy the condition $m_{1}+m_{2}+\cdot\cdot\cdot+m_{k}=k$ and $m_{1}+2\,m_{2}+\cdot\cdot+k\,m_{k}=k$ would be $m_{1}=k$ and $m_{l}=0\,\forall\,l\neq1$. This gives,	
\begin{equation}
\partial_{\theta}^{k}U=\partial_{z}^{k}U\,\mathcal{O}\left(\bar{h}^{k}\right),
\end{equation}
Also, from $\left(\ref{U_partial_z}\right)$, the order of $\partial_{z}^{k}U=\mathcal{O}\left(2^{k}\right)$, which gives,
\begin{equation}
\label{Order_U}
\partial_{\theta}^{k}U=\mathcal{O}\left(2^{k}\bar{h}^{k}\right),
\end{equation}
Lastly, as $\frac{1.3.5\cdot\cdot\cdot k-1}{\left(\frac{k}{2}!\right)^{2}\left(\frac{k}{2}-1\right)!}<1\;\forall k\,\,$\footnote{Let $P\left(k\right)=\frac{1.3.5\cdot\cdot\cdot k-1}{\left(\frac{k}{2}!\right)^{2}\left(\frac{k}{2}-1\right)!}$, where $k=4,6\cdot\cdot$ then, using mathematical induction, 
\begin{equation}
P\left(4\right)=\frac{3}{4}<1
\end{equation}
Let $P\left(k\right)<1$ holds true, 
\begin{equation}
\begin{split}
P\left(k+2\right)&=\frac{1.3.5\cdot\cdot\cdot k-1\,k+1}{\left(\frac{k+2}{2}!\right)^{2}\left(\frac{k+2}{2}-1\right)!}=\frac{1.3.5\cdot\cdot\cdot k-1\,k+1}{\left(\frac{k}{2}+1\right)^{2}\left(\frac{k}{2}!\right)^{2}\left(\frac{k}{2}\right)\left(\frac{k}{2}-1\right)!}\\&=P\left(k\right)\frac{k+1}{\left(\frac{k}{2}+1\right)^{2}\left(\frac{k}{2}\right)}=P\left(k\right)\,\frac{1}{k\left(k+1\right)+2k+\frac{k}{k+1}}<1\,\forall k
\end{split},
\end{equation}
Therefore, the statement $P\left(k\right)<1$ holds true.
}, then using $\left(\ref{Order_U}\right)$, the order of the $k^{th}$ term can be obtained as,
\begin{equation}
\epsilon\,\partial_{\theta}^{k}U\left(\theta_{0}\right)A_{w}^{k-2}\frac{1}{2^{k-2}}\frac{1.3.5\cdot\cdot\cdot k-1}{\left(\frac{k}{2}!\right)^{2}\left(\frac{k}{2}-1\right)!}\frac{1}{N^{\frac{k}{2}-1}}=\epsilon\,\mathcal{O}\left(\frac{A_{w}^{k-2}\bar{h}^{k}}{N^{\frac{k-2}{2}}}\right),
\end{equation}
Now, if $\frac{A_{w}\bar{h}}{\sqrt{N}}<1$,then $3^{rd},\,4^{rth}\cdot\cdot\cdot$ terms of $\left(\ref{perturbed_freq_expansion_appendix}\right)$ can be written as,
\begin{equation}
\mathcal{O}\left(\frac{A_{w}^{2}\bar{h}^{4}}{N}\right)+\mathcal{O}\left(\frac{A_{w}^{4}\bar{h}^{6}}{N^{2}}\right)+\cdot\cdot\cdot=\mathcal{O}\left(\frac{A_{w}^{2}\bar{h}^{2}}{N}\right)+\mathcal{O}\left(\frac{A_{w}^{4}\bar{h}^{4}}{N^{2}}\right)+\cdot\cdot\cdot=\mathcal{O}\left(\frac{A_{w}^{2}\bar{h}^{2}}{N}\right),
\end{equation}
Using the property, $\mathcal{O}\left(c\,g\right)=\mathcal{O}\left(g\right)$, where, $c$ is a constant and $\mathrm{{if,}}f_{1}=\mathcal{O}\left(g_{1}\right)\,and\,f_{2}=\mathcal{O}\left(g_{2}\right)$, then, $f_{1}+f_{2}=\mathcal{O}\left(max\left(g_{1},g_{2}\right)\right)$.
Finally, using the above result, $\left(\ref{perturbed_freq_expansion_appendix}\right)$ can be written as,
\begin{equation}
\begin{split}
\Omega_{w}^{2}&=\alpha_{w}^{2}+\epsilon\,2a^{2}x_{0}\left(2h-x_{0}\right)+\epsilon\,\mathcal{O}\left(\frac{A_{w}^{2}\bar{h}^{2}}{N}\right)\\&=\alpha_{w}^{2}+\epsilon\,2a^{2}x_{0}\left(2h-x_{0}\right)+\epsilon\,\mathcal{O}\left(a^{2}h^{2}\frac{E_{w}}{N}\right)
\end{split},
\end{equation}
where, $A_{w}=\sqrt{2E_{w}}$ and $\bar{h}=ah$.

\end{document}